\documentclass[journal]{IEEEtran}

\usepackage{cite}
\usepackage{amsmath,amssymb,amsfonts}
\usepackage{algorithm}
\usepackage{algpseudocode}
\usepackage{graphicx}
\usepackage{textcomp}
\usepackage{xcolor}
\usepackage{tabularx}
\newtheorem{Theorem}{Theorem}
\newtheorem{Lemma}{Lemma}

\newtheorem{Definition}{Definition}

\newtheorem{Remark}{Remark}
\usepackage{graphicx}
\usepackage{array}
\usepackage{caption}
\usepackage{subcaption}
\usepackage{adjustbox}
\usepackage[utf8]{inputenc}
\usepackage{mathtools}
\usepackage{balance}

\def\Ac{{\mathcal A}}

\def\Abf{{\mathbf A}}

\def\Bc{{\mathcal B}}

\def\Bbf{{\mathbf B}}

\def\Cc{{\mathcal C}}

\def\Cbf{{\mathbf C}}

\def\Dc{{\mathcal D}}

\def\Ec{{\mathcal E}}

\def\Fc{{\mathcal F}}

\def\Ic{{\mathcal I}}

\def\Lbf{{\mathbf L}}

\def\Mc{{\mathcal M}}

\def\Nbb{{\mathbb N}}

\def\Qc{{\mathcal Q}}

\def\Rbb{{\mathbb R}}

\def\Sbb{{\mathbb S}}

\def\Tc{{\mathcal T}}

\def\Uc{{\mathcal U}}

\def\Ubf{{\mathbf U}}
\def\ubf{{\mathbf u}}

\def\Wc{{\mathcal W}}

\def\Xc{{\mathcal X}}

\def\Zbb{{\mathbb Z}}

\def\0{{\bf 0}}

\def\underdotline#1{{\setbox0=\hbox{#1}\rlap{\raise-0.28em\hbox to\wd0{\rm\tiny\cleaders\hbox{.\kern-0.1ex}\hfill}}\box0}}
\def\underdashline#1{{\setbox0=\hbox{#1}\rlap{\raise-0.28em\hbox to\wd0{\rm\tiny\cleaders\hbox{-\kern-0.1ex}\hfill}}\box0}}
\def\understarline#1{{\setbox0=\hbox{#1}\rlap{\raise-0.28em\hbox to\wd0{\rm\tiny\cleaders\hbox{$\star$\kern-0.1ex}\hfill}}\box0}}

\newcommand{\bitem}{\begin{itemize}}
\newcommand{\eitem}{\end{itemize}}
\newcommand{\btabular}{\begin{tabular}}
\newcommand{\etabular}{\end{tabular}}
\newcommand{\bcenter}{\begin{center}}
\newcommand{\ecenter}{\end{center}}
\newcommand{\bea}{\begin{eqnarray}}
\newcommand{\eea}{\end{eqnarray}}
\newcommand{\bean}{\begin{eqnarray*}}
\newcommand{\eean}{\end{eqnarray*}}

\newcommand{\ba}{\left. \begin{array}}
\newcommand{\ea}{\\ \end{array} \right.}
\newcommand{\bab}{\left[ \begin{array}}
\newcommand{\eab}{\\ \end{array} \right]}
\newcommand{\bap}{\left( \begin{array}}
\newcommand{\eap}{\\ \end{array} \right)}
\newcommand{\bbm}{ \begin{bmatrix}}
\newcommand{\ebm}{\\ \end{bmatrix} }

\newcommand{\bear}{\begin{array}}
\newcommand{\eear}{\\ \end{array}}

\newcommand{\ovl}{\overline}

\newcommand{\td}{\tilde}

\newcommand{\non}{\nonumber}

\newcommand{\ra}{\rightarrow}

\def\Qced{{\ \vrule width 1.5mm height 1.5mm \smallskip}}

\font\myownfont=cmr17 scaled \magstep5
\def\psfancypar#1#2{\def\biginitial#1{{\myownfont#1}}%
  \def\makeinitial#1{\setbox8\hbox{\strut\vbox to 1.3ex
    {\hbox{\biginitial#1}\vskip -4pc plus 3.5pc minus 3.5pc}}}%
  \makeinitial#1%
  \ifdim\parindent>1.3\wd8\dimen8=\parindent
     \else\dimen8=1.3\wd8\fi
  \hangindent=\dimen8\hangafter=-2
  \noindent
  \strut\hskip-1\dimen8\box8{\sc#2}}%

\newcounter{subequation}
\def\beasub{\addtocounter{equation}{+1}
\setcounter{subequation}{\value{equation}}
\setcounter{equation}{0}
\renewcommand{\theequation}{\arabic{subequation}\alph{equation}}
\begin{eqnarray}}
\def\eeasub{\end{eqnarray}
\setcounter{equation}{\value{subequation}}
\renewcommand{\theequation}{\arabic{equation}}}

\begin{document}

\title{
Encrypted Observer-based Control for Linear Continuous-Time Systems
}

 \author{Hung Nguyen$^{1}$, Binh Nguyen$^{2}$, Hyung-Gohn Lee$^{1}$ and  Hyo-Sung Ahn$^{1}$, {\it Senior Member,~IEEE}
 \thanks{$^1$Hung M. Nguyen, Hyung-Gohn Lee and Hyo-Sung Ahn are with the School of Mechanical Engineering, Gwangju Institute of Science and Technology (GIST), Gwangju 61005, South Korea (e-mail: nguyenmanhhung@gist.ac.kr; hyunggohnlee@gm.gist.ac.kr;  hyosung@gist.ac.kr).}
 \thanks{$^2$Binh T. Nguyen is with College of Engineering, Texas A\&M University-Corpus Christi,  Corpus Christi, TX 78412, United States (email: binh.nguyen@tamucc.edu).}
 }
\maketitle

\begin{abstract}
This paper is concerned with the stability analysis of 
encrypted observer-based control for
linear continuous-time systems.
Since conventional encryption has limited ability to deploy in continuous-time integral computation, our work presents systematically a new design of encryption for a continuous-time observer-based control scheme.
To be specific, in this paper, both control parameters and signals are encrypted by the learning-with-errors (LWE) encryption to avoid data eavesdropping.
Furthermore, we propose encrypted computations for the observer-based controller based on its discrete-time model,
and present a continuous-time virtual dynamics of the controller for further stability analysis. 
Accordingly, we present novel stability criteria
by introducing linear matrix inequalities (LMIs)-based conditions associated with quantization gains and sampling intervals.
The established stability criteria with theoretical proofs based on a discontinuous Lyapunov functional possibly provide a way to select quantization gains and sampling intervals to guarantee the stability of the closed-loop system.
Numerical results on DC motor control corresponding to several quantization gains and sampling intervals demonstrate the validity of our method.
\end{abstract}


\section{Introduction}
In recent years, the development of cloud computing has received great attention in many modern control systems such as smart grids, intelligent transportation systems, and robotics \cite{zijian2017optimal,shengdong2019intelligent,yin2022cloud,wang2016hierarchical,jiang2022multi}, 
and the security threat has been one of the main issues \cite{he2021secure,teixeira2015secure,mousavinejad2018novel}. 
There are many types of attacks, such as Denial of Service (DoS) attack \cite{yang2020distributed, liu2020secure,zhu2021novel}, zero dynamic attack \cite{teixeira2015secure}, reply attack \cite{mousavinejad2018novel} and eavesdropping attack \cite{darup2021encrypted}. Among these attacks, eavesdropping is a basic one, which is performed to steal confidential information to apply more advanced attacks. 

\begin{figure*}
    \centering
    \includegraphics[width = \linewidth]{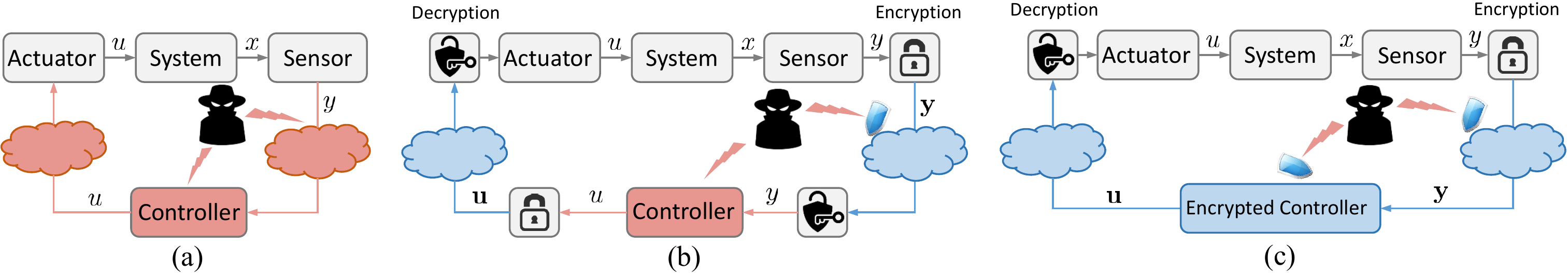}
    \caption{(a) A cloud-based control scheme without encryption, (b) a cloud-based control scheme with encryption-based secure communications and an unencrypted controller, and (c) a cloud-based control scheme with both encryption-based secure communications and controller. The red indicates the parts of the system vulnerable to the attack, and the blue represents the parts protected against the attack by encryption.} \label{cloud-based}
\end{figure*}

In eavesdropping attacks, communicating and processing data on a third-party platform may lead to data eavesdropping. As can be seen from Fig. \ref{cloud-based}, an eavesdropper is able to steal data through both communication and collaboration with the controller without encryption.
Additionally, even if the communications are encrypted as in Fig. \ref{cloud-based}b, the attacker can still collaborate with the controller for data eavesdropping.
Harnessing a homomorphic encryption (HE), an encrypted controller can perform directly on encrypted signals and parameters without decryption (Fig. \ref{cloud-based}c), which could protect the control system from eavesdroppers \cite{darup2021encrypted,teranishi2019stability,murguia2020secure,kogiso2015cyber,fujita2015security,farokhi2016secure}.
A model predictive control was implemented using additively HE in \cite{schulze2020encrypted} and \cite{darup2017towards}.
A state-vector estimator using a private Extended Kalman Filter was proposed in \cite{gonzalez2014state}.
Secure distributed control schemes based on encryption for multi-agent systems were considered in \cite{ruan2019secure,gao2021encryption,fang2021secure,gao2021fault}.
The work \cite{farokhi2017secure} applied the Paillier encryption \cite{paillier1999public} to encrypt the control signals for a linear control system, and a static output feedback controller was considered. 
In \cite{kim2022dynamic}, the authors provided a dynamic feedback controller over encrypted data utilizing homomorphic features of cryptosystems, whose performance is similar to the linear dynamic controllers over real-valued data. By applying an integer conversion for the state matrix of the controller without scaling, they showed that a system could be converted to another system with the same input-output relation. However, the integer conversion process can only be applied to a class of systems under certain conditions. In \cite{teranishi2020dynamic}, the authors proposed a quantization design for a linear control system, in which not only the controller parameters are quantized, but the control signals are also quantized. 
Using a dynamic quantizer with a sensitivity that depends on the system's state and control signals, they showed that asymptotic stability could be achieved.  

However, there have still been many challenging issues in the encrypted control systems.
Firstly, the encrypted controller can only perform with integers \cite{teranishi2020dynamic, kim2022dynamic}, which requires a quantization process that results in quantization errors.
For dynamic controllers such as observer-based controllers, the quantization errors could be accumulated over time if the state matrix of the controller contains at least one non-integer element \cite{kim2022dynamic}.
Secondly, since plants are continuous-time while the encrypted controllers are formulated in terms of integer computation, that leads to difficulties in the stability analysis of the encrypted control systems in the presence of the quantization errors and sampling-data process.
Both quantization gains and sampling intervals have critical impacts on the encrypted observer-based control system analysis.
As far as we are concerned, there has been little progress toward the criteria of quantizer and sampling intervals for encrypted observer-based continuous linear time-invariant (LTI) systems.

To address these difficulties, this paper investigates the design of the quantizers and sampling intervals for the encrypted observer-based control systems. 
The contributions of this work are summarized as follows.
\begin{itemize}
    \item [1)] First, we make the first attempt to consider an encrypted observer-based controller for a continuous-time LTI system. 
    Since encryption does not support continuous-time integral, the observer-based controller should be formulated in the discrete-time form. However, the plant is continuous-time, which results in difficulties in the stability analysis of the encrypted control system.
    Taking advantage of discretizing continuous-time Luenberger observers,
    this paper presents a novel encrypted observer-based controller allowing us to analyze the continuous-time stability of the closed-loop system.
    By introducing continuous-time virtual dynamics of the encrypted controller, we succeed in formulating the closed-loop system in 
    a sampled-data system and providing stability criteria for the system.
    Unlike many existing works \cite{teranishi2020dynamic,teranishi2019stability,murguia2020secure,kogiso2015cyber}, where the stability analysis is concerned with discrete-time systems, 
    we provide a continuous-time stability analysis taking into account both the quantization and sampling actions.
    \item[2)] 
    Differing from \cite{teranishi2020dynamic,teranishi2021encrypted}, our work presents conditions for the selection of quantization gains and sampling intervals by which all quantization gains can be predetermined independently from system signals (e.g., system outputs and estimated states).
    It is worth noting that the determination of the quantization gains (or sensitivity) from \cite{teranishi2020dynamic,teranishi2021encrypted} is required at each time step and also needs the system signals.
    \item[3)] Finally, we introduce novel stability criteria in terms of LMIs-based conditions and give theoretical proof based on a discontinuous Lyapunov functional for sampled-data systems.
    We also provide some novel results for stability analysis of a linear sampled-data system with disturbances by introducing a framework in which global asymptotic stability of a linear sampled-data system can be ensured under bounded energy disturbance.
    The proposed criteria are associated with the quantization gains and sampling interval, and by selecting appropriate values for the quantization gains and sampling interval, the global asymptotic stability of the closed-loop system can be ensured.
\end{itemize}

\textbf{Notations:}
The set of real numbers, integers, non-negative real numbers, positive real numbers, positive integers, and natural numbers are denoted by $\Rbb, \Zbb, \Rbb_+, \Rbb_{++}, \Zbb_{++},\Nbb$, respectively. 
The symbol $\Rbb^{n \times m}$ indicates the set of matrices with $n$ rows and $m$ columns. The set of symmetric matrices with dimension $n$ is denoted by $\Sbb^{n}$. Additionally, $\Sbb^{n}_{++}$ represents the set of positive definite symmetric matrices. 
Describe the set of modulo $q \in \mathbb{N}$ by $\mathbb{Z}_q = \{0,1,...,q-1\}$. The symbol $\lfloor \cdot \rceil$ denotes the rounding function. For a vector, $\|\cdot\|$ indicates the Euclidean norm, while for a matrix, $\|\cdot\|$ represents the Frobenius norm. For a symmetric matrix, its minimum and maximum eigenvalues are denoted by $\lambda_{min}(\cdot)$ and $\lambda_{max}(\cdot)$, respectively. 
For any matrix $\Xc$, the notations $\Xc \succ 0$ (or $\Xc \succeq 0$) and $\Xc \prec 0$ (or $\Xc \preceq 0$) mean that $\Xc$ is a positive definite (or positive semi-definite) matrix and a negative definite (or negative semi-definite) matrix, respectively. 
The symbols $I$ and $0$ represent the identity and zero matrices with suitable dimensions, respectively; additionally, $I_n$ represents the identity matrix with $n$ rows and $n$ columns. For any entity $z$, $z(t_k)$ denotes the value of $z$ during $[t_k, t_{k+1})$. In symmetric block matrices, the asterisk $(\ast)$ indicates the terms induced by symmetry. 
For any square matrix $\Qc$, $\mathbf{He} \{\Qc\} = \Qc + \Qc^\top$. The notation $\times$ represents the multiplication of two matrices or the Cartesian product of two sets.

\section{Preliminaries} \label{preliminaries}

\subsection{Encryption}
This subsection introduces the cryptosystem utilized in this paper. For an unencrypted value, we call it a plaintext while an encrypted value is called a ciphertext. Denote $\mathcal{C}$ the ciphertext space, let ${\rm{Enc}}: \mathbb{Z}_q \rightarrow \mathcal{C}$ represent the encryption map, and ${\rm{Dec}}: \mathcal{C} \rightarrow \mathbb{Z}_q$ denote the decryption map. We briefly describe LWE-based encryption \cite{kim2020comprehensive} as follows.
\begin{itemize}
\item Choose a private key $k_p \in \mathbb{Z}_q^{n}$, a random vector $a \in \mathbb{Z}_q^n$, and a small random error $e \in \mathbb{Z}_q$.
\item \textbf{Encryption:} For a plaintext $m \in \mathbb{Z}_q$, the corresponding ciphertext is computed as  
\begin{align*}
    {\rm{Enc}}(m) \triangleq  \left(\begin{bmatrix}
                       m + k_p^\top a + e\\
                          a
                     \end{bmatrix}~~{\rm{mod}} \; q \right) = c \in \Zbb_q^{n+1}. 
\end{align*} 

\item \textbf{Decryption:} For the ciphertext $c \in \Cc$ corresponding to the plantext $m \in \Zbb_q$, define
\begin{align*}
    {\rm{Dec}}(c) \triangleq \begin{bmatrix}
                    1& -k_p^\top 
                    \end{bmatrix} c ~~{\rm{mod}} \; q.
\end{align*}
It is obvious that ${\rm{Dec}}(c) = m + e$. To deal with the error $e$, let us consider the encryption with gain $G>0$ as
\begin{align}
   c_G = {\rm{Enc}}(Gm) = \begin{bmatrix}
                       Gm + k_p^\top a + e\\
                          a
                     \end{bmatrix}~~{\rm{mod }} \; q. \label{G_gain}
\end{align}
Then, with $G \in \mathbb{Z}_q$ being a positive gain such that $G \!>\!2e$, the  plaintext can be recovered as
\begin{align*}
  \left\lfloor\frac{{\rm{Dec}}(c_G)}{G}\right\rceil = \left\lfloor m + \frac{e}{G}  \right\rceil = m. 
\end{align*}
\end{itemize}
Hereafter, when considering a ciphertext, we always assume that it is encrypted with a suitable gain $G$.

\subsubsection{Additive property} For $c_1, c_2 \in \mathcal{C}$, $c = c_1 + c_2~~{\rm{mod}} \; q$, and $c^\prime = k c_1 ~ {\rm{mod}} \; q$ with $k\in\mathbb{Z}_q$, one has
\begin{align*}
&{\rm{Dec}}(c) = {\rm{Dec}}(c_1) + {\rm{Dec}}(c_2), \\
&{\rm{Dec}}(c^\prime) = k  {\rm{Dec}}(c_1).
\end{align*}

\subsubsection{Multiplication}
To have the ability of multiplication, let us introduce a separate algorithm for encrypting the multipliers utilized in \cite{gentry2013homomorphic}. 
Consider the ciphertexts $c_1, c_2 \in \Zbb_q^{n+1}$ corresponding to the plaintexts $m_1, m_2 \in \Zbb_q$. Choose $q \in \Zbb_{++}$ such that there exist $\omega, d \in \Zbb_q$ and $\omega^d = q$.
Let $\Dc(\cdot)$ denote the function that decomposes the argument by its string of digits as $\Dc(b) = [b_1,\dots,b_{d}]^\top$ with $0 \le b_i \le \omega-1$, $\forall i \in \{1,2,\dots,d\}$, and $b\in\Zbb_q$. 
Thus, the ciphertext $c_2$ could be written as
\begin{align}
    c_2 = H  \Dc(c_2), \label{decompostition}
\end{align}
with $H = \bbm I_{n+1} & \omega I_{n+1} & \dots & \omega^{d-1} I_{n+1}\ebm$. To illustrate the decomposition in \eqref{decompostition}, let us consider an example with $q = 16$, $\omega = 2$ and $d = 4$, it holds $\omega^d = q$. Then, the integer $c_2 = 13$ can be written as $c_2 = H \Dc(c_2)$ with $H = \bbm 2^0 & 2^1 & 2^2 & 2^3 \ebm$ and $\Dc(c_2) = \bbm 1 & 0 & 1 & 1 \ebm^{\top}$. The multiplier is then encrypted as follows
\begin{align}
    {\rm{Enc}^\prime}(m_1) &= m_1 H + \bbm k_p a_1 & k_p a_2 & \dots & k_p a_{d(n+1)} \\ a_1 & a_2 & \dots & a_{d(n+1)}\ebm \nonumber \\
    &~+ \bbm e_1 & e_2 & \dots & e_{d(n+1)} \\ {0}_n & {0}_n & \dots & {0}_n\ebm~~{\rm{mod}} ~ q.
    \label{GSW}
\end{align}
Then, the multiplication between two ciphertexts $c_1$ and $c_2$ is defined as
\begin{align*}
    c_1 \odot c_2  \triangleq {\rm{Enc}}^\prime(m_1) \Dc(c_2) ~~ {\rm{mod}} ~ q.
\end{align*}
To see the homomorphic property, we note that 
\begin{align}
    {\rm{Dec}}({\rm{Enc}}^\prime(m_1) \Dc(c_2)) = m_1 m_2 + e ~~ {\rm{mod}} \; q, \label{c1}  
\end{align}
with $e = \bbm e_1 & \dots & e_{d(n+1)} \ebm \Dc(c_2)$. To deal with the error in \eqref{c1}, we similarly consider the encryption with a gain $G$ as in \eqref{G_gain}, and by selecting a large enough value of $G$, the error in \eqref{c1} vanishes and the multiplication $m_1 m_2$ could be exactly recovered.
Note that the encryption and decryption functions can be applied element-wisely to vectors or matrices.
\subsection{Quantization}
Consider the uniform quantizer with the following form 
\begin{align}
    Q_{\Theta}(x) = \frac{1}{\Theta}\left( \lfloor \Theta x \rceil \right), \label{b1}
\end{align}
where the quantization gain $\Theta >0$ is a positive value. For the uniform quantizer \eqref{b1}, with $x \in \Rbb^n$ and $X \in \Rbb^{m\times n}$, the bounds on the quantization errors are always given as 
\begin{align}
    \| Q_{\Theta}(x) - x \| \le \frac{\sqrt{n}}{2\Theta},~~ \| Q_{\Theta}(X) - X \| \le \frac{\sqrt{mn}}{2\Theta}.
    \label{quantization_error_bound}
\end{align}
We note that if the quantization gain $\Theta$ is a fixed value, then the quantizer is static. However, if the quantization gain varies over time, the quantizer is called a dynamic quantizer.

\subsection{Sampled-data observer-based controller}
For a linear continuous-time system
\begin{align}
\left\{ \begin{array}{l}
\dot x(t) = {A}x(t) + {B}u(t),\\
y(t) = {C}x(t),
\end{array} \right. \label{a1}
\end{align} 
let $x(t) \in \mathbb{R}^n$, $u(t) \in \mathbb{R}^m$ and $y(t) \in \mathbb{R}^r$ stand for the state vector, control input, and output, respectively. 
The system's matrices  $A, B$, and $C$ are given in appropriate dimensions.
We consider a sampled-data observer-based controller to stabilize the system \eqref{a1}.
Since the system \eqref{a1} is continuous-time, we take the set of sampling instants $\mathcal{I} = \{t_1,\dots,t_i,\dots\}$, with $\lim_{i\rightarrow \infty} t_i = \infty$, and the constant time interval $h = t_{i+1} - t_{i} > 0, \forall i \ge 0$.  
Let $\chi(t)$ be the estimation of $x(t)$ and assume that the matrices $L, C, K$ are given in appropriate dimensions. Then the conventional Luenberger's observer-based controller is given in the following form
\begin{subequations} \label{obsv}
\begin{align}
\dot{\chi}(t) &= A \chi(t) + B u(t) + L(y(t_k)-C\chi(t_k)),~
\label{Obs_cal_uncryp}
\\
u(t) &= K \chi(t_k), \quad \forall t \in [t_k, t_{k+1}).
\label{Ctrl_cal_uncryp}
\end{align} \label{Obs_Ctrl_uncryp}
\end{subequations}
To investigate the stability of the sampled-data observer-based control system (\ref{a1}) and (\ref{obsv}), consider the following linear sampled-data system 
\begin{align}
    \dot{z}(t) = \mathcal{A}z(t) + \mathcal{A}_c z(t_k) + \eta(t), \label{sd_sys:org}
\end{align}
where $\mathcal{A}$, $\mathcal{A}_c$ are given constant matrices, and $\eta(t)$ is the disturbance.

\begin{Definition}[Integral Quadratic Constraint \cite{fetzer2016general}] \label{Def1}
The system (\ref{sd_sys:org}) is said to satisfy integral quadratic constraint (IQC) if for any initial condition $z_0 = z(t_0)$, its solution satisfies 
\begin{align}
\lim_{t \ra \infty}  \int_{t_0}^{t}  \| z(\tau) \|^2 d\tau < \infty.
\label{sampledIQC}
\end{align}
\end{Definition}
The IQC in Definition \ref{Def1} also implies global asymptotic stability of (\ref{sd_sys:org}).
The following lemmas are useful for stability analysis of the sampled-data system (\ref{sd_sys:org}).
\begin{Lemma} \label{Lemma_LMIs_with_Uncertainty}
Let $\Gamma = \Gamma^\top, \Pi_1, \Pi_2, \Omega$ and $\Delta$ be the matrices with appropriate dimensions. Then, the inequality
$\Gamma + \Pi_1 \Delta^\top \Pi_2 + \Pi_2^\top \Delta \Pi_1^\top + \Delta \Omega \Delta^\top \preceq 0$ holds if
there exist $\epsilon, \kappa >0$ such that
\begin{subequations}
\begin{align} 
\bbm 
\Gamma + \kappa (\Pi_1 \Pi_1^\top+I) & \Pi_2^\top\\
\Pi_2 & -\epsilon I 
\ebm \preceq 0, \label{condi_a_lemma_3}
\\
\Omega - \epsilon I \preceq 0. \label{condi_b_lemma_3}
\end{align} \label{condi_lemma_3}
\end{subequations}
\end{Lemma}
\textit{Proof:} The proof can be completed using $\mathcal{S}$-procedure \cite{boyd1994linear}.

\begin{Lemma} \cite{gyurkovics2015note} \label{Lemma_integral_inequality}
Let $R \succ 0$, and $z(t)$ be a differential function. Then for all matrices $N_1, N_2$ given in appropriate dimensions, the following inequality holds
\begin{align}
    \int_{t_k}^{t} \dot{z}^\top(\tau) R \dot{z}(\tau) d\tau  \ge \xi^\top(t) \Phi \xi(t),
\end{align}
where $\Phi = \mathbf{He}\big\{\Phi_1^\top N_1 + \Phi_2^\top N_2 \big\} - (t-t_k) 
    \Big( N_1^\top R^{-1} N_1 + \frac{1}{3} N_2^\top R^{-1} N_2 \Big), \Phi_1 = E_1 - E_2, \Phi_2 = E_1 + E_2 -2E_3,E_1 = \bbm I&0&0 \ebm, E_2 = \bbm 0&I&0 \ebm, E_3 = \bbm 0&0&I \ebm$ and $\xi(t) = [z^\top(t),z^\top(t_k), \frac{1}{t-t_k} \int_{t_k}^t z^\top(\tau) d\tau]^\top$.
\end{Lemma}

\section{Proposed Encrypted Observer-based Control} \label{Prob_formulation}
In this section, we first propose an encrypted observer-based controller for a continuous LTI system in which, the control parameters and signals are encrypted to avoid eavesdropping attacks. Second, we formulate the closed-loop system in the form of a linear sampled-data system with uncertainties and disturbance.



\subsection{Observer-based secure control scheme}
\begin{figure}[t]
    \centering
    \includegraphics[width = \linewidth]{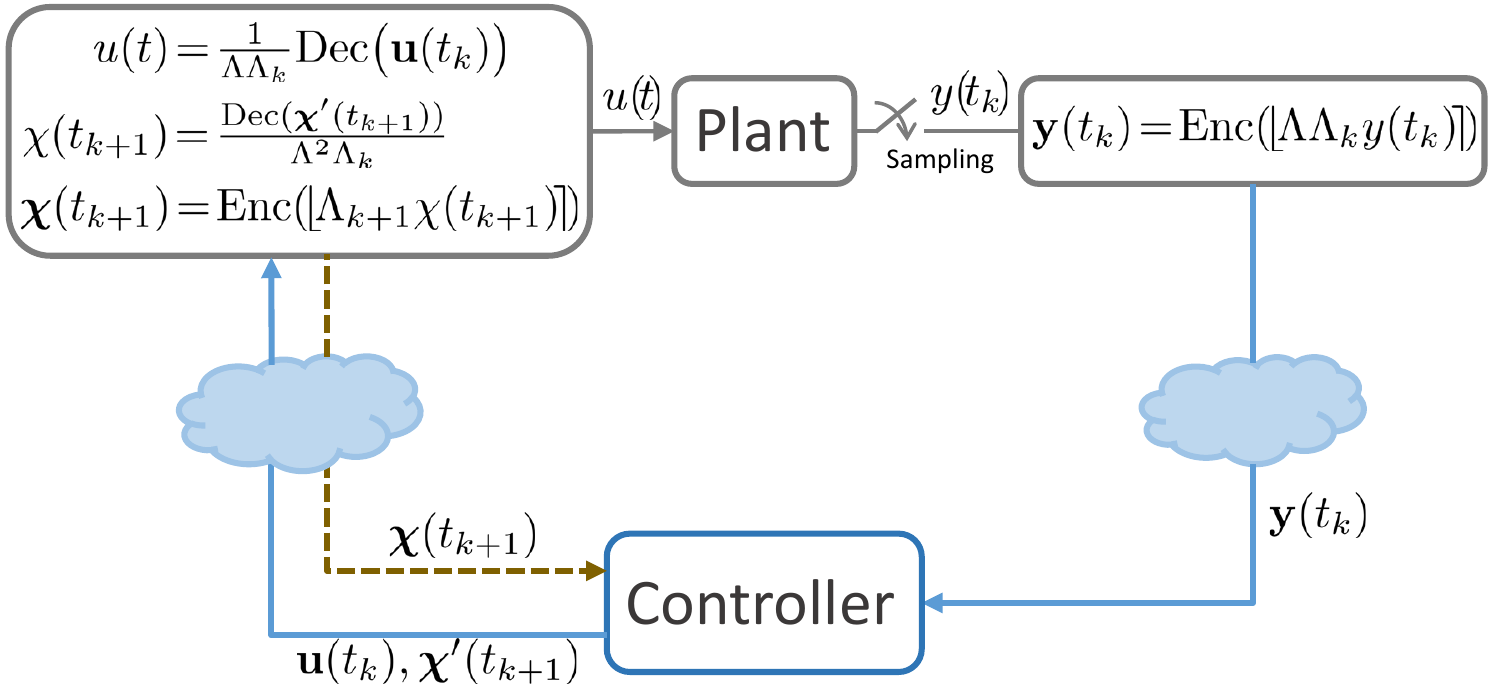}
    \caption{Encrypted control system diagram }
    \label{Fig_control_diagram}
\end{figure}

In this part, we present a secure version of the observer-based controller \eqref{Obs_Ctrl_uncryp}, in which the objective is to stabilize the system \eqref{a1} while ensuring security with the use of LWE-based encryption. For this purpose, all parameters and control signals in \eqref{Obs_Ctrl_uncryp} need to be encrypted. It can be seen that \eqref{Obs_cal_uncryp} cannot be computed based on encrypted signals since LWE-based encryption does not allow computing on continuous-time integral.
Thus, in order to compute the observer-based controller \eqref{Obs_cal_uncryp} in an encrypted way, we first take advantage of its solution at each sampling time as follows
\begin{subequations}
\begin{align}
    \chi(t_{k+1}) &= A_d \chi(t_k) \!+\! B_d u(t_k) \!+\! L_d \left( y(t_k) \!-\! C \chi(t_k) \right) ,
    \label{dis_uncryp}
    \\ 
    u(t_k) &= K \chi(t_k),
\end{align} \label{dis_ctrl_uncryp}
\end{subequations}
with $A_d = e^{Ah}, B_d = \int_0^h e^{A\tau} B d\tau$ and $L_d = \int_0^h e^{A\tau} L d\tau$. 
Given that the controller \eqref{dis_ctrl_uncryp} is executed at the computational unit, which is possibly located far away from the plant, then for security purposes, the unencrypted matrices  $A_d, B_d, L_d, C, K$ and values $y(t_k)$ are not sent directly to the computational unit. Instead,
the following encrypted values obtained through (\ref{G_gain}) and (\ref{GSW}) based on LWE are utilized
\begin{align*}
    &\mathbf{y}(t_k) = {\rm{Enc}}\left(\left\lfloor \Lambda \Lambda_k{y(t_k)} \right\rceil\right), 
    \mathbf{A}_d = {\rm{Enc}^\prime}\left(\left\lfloor \Lambda^2 {A_d}\right\rceil\right), 
    \\
    &\mathbf{B}_d = {\rm{Enc}^\prime}\left(\left\lfloor \Lambda {B_d} \right\rceil\right), 
    \Lbf_d =  {\rm{Enc}^\prime}\left(\left\lfloor \Lambda {L_d} \right\rceil\right), 
    \\
    &\Cbf = {\rm{Enc}^\prime}\left(\left\lfloor \Lambda {C} \right\rceil\right),~
    \mathbf{K} = {\rm{Enc}^\prime}\left(\left\lfloor \Lambda {K} \right\rceil\right),
\end{align*}
with a positive value $\Lambda$ and a possibly time-varying positive value $\Lambda_k$.
It is worth mentioning that, similar to \cite{kim2022dynamic}, we use different quantization gains for different entities.  
The quantizers with static gains are employed to quantize the matrices $A_d, B_d, L_d, C$ and $K$ in the controller \eqref{dis_ctrl_uncryp}, while the quantizer with a possible dynamic gain is utilized for quantizing the output measurement $y(t_k)$.
By letting $\boldsymbol{\chi}(t_k) = {\rm{Enc}}\left( \left \lfloor \Lambda_k {{\chi}(t_k)}  \right \rceil \right)$ and $\boldsymbol{\chi}^\prime(t_k) = {\rm{Enc}}\left( \left \lfloor \Lambda^2 \Lambda_{k-1}{{\chi}(t_k)}  \right \rceil \right)$, \eqref{dis_ctrl_uncryp} is encrypted as
\begin{subequations}
\begin{align}
    \boldsymbol{\chi}^\prime(t_{k+1}) &= \Abf_d \Dc \big(\boldsymbol{\chi}(t_k)\big) + \Bbf_d \Dc\big( \mathbf{u}(t_k) \big)  \nonumber  \label{update_chi_encrypted}\\
    & \quad + \mathbf{L}_d \left( \Dc\big(\mathbf{y}(t_k)\big) - \Dc\big(\Cbf \Dc\big(\boldsymbol{\chi}(t_k)\big)\big) \right),
    \\
    \mathbf{u}(t_k) &= \mathbf{K} \Dc\big(\boldsymbol{\chi}(t_k)\big). \label{update_u_encrypted}
\end{align} \label{enc_controller}
\end{subequations}
Here, we deploy a similar approach to that described in \cite{kim2022dynamic} to release the control input $u(t)$ from the encrypted controller \eqref{enc_controller} (see Fig. \ref{Fig_control_diagram}). 
The proposed controller \eqref{enc_controller} is performed over encrypted data to release encrypted control input $\ubf(t_k)$. 
Then, the control input $u(t)$ for the plant \eqref{a1} can be obtained by the decryption
$u(t) = \frac{1}{\Lambda \Lambda_k}{\rm{Dec}(\mathbf{u}(t_k))}, \forall t \in [t_k, t_{k+1}).$
In the update \eqref{update_chi_encrypted}, the encrypted value $\boldsymbol{\chi}^\prime(t_{k+1})$ is computed at time slot $t_k$; however, it is not used to compute the encrypted control input \eqref{update_u_encrypted} at time $t_{k+1}$ $\left(\boldsymbol{\chi}^\prime(t_{k+1}) \neq \boldsymbol{\chi}(t_{k+1})\right)$.
Thus, besides sending $\mathbf{u}(t_k)$, the controller also sends $\boldsymbol{\chi}^\prime(t_{k+1})$ to the plant at time slot $t_k$.
Then, the encrypted value $\boldsymbol{\chi}(t_{k+1})$ can be computed and sent to the controller by the plant as
\begin{align*}
    \chi(t_{k+1}) &= \frac{{\rm{Dec}}(\boldsymbol{\chi}^{\prime}(t_{k+1}))}{\Lambda^2 \Lambda_k},
    \\
    \boldsymbol{\chi}(t_{k+1}) &= {\rm{Enc}}\left( \left\lfloor \Lambda_{k+1} \chi(t_{k+1})  \right\rceil \right).
\end{align*}
Consequently, the controller only holds the encrypted values of the system parameters and control signals; thus, the security of the control system is guaranteed.
%

\subsection{Linear sampled-data system formulation}
We note here that a quantization process for encryption is required; further, when applying to any value, the encryption-decryption procedure generates the same result as the quantization process.
Therefore, in terms of stability analysis, the implementation of LWE-based encryption to the observer-based controller \eqref{enc_controller} is equivalent to the one obtained with only quantization as follows
\begin{align}
     \begin{cases}
        {\chi}_d(t_{k+1}) \!\!\!\!\!\!&= \bar{A}_d \bar{\chi}_d(t_k)  + \bar{B}_d u(t_k) 
        \\
        \!\!\!\!\!\!&\hspace{1.85cm} + \bar{L}_d (\bar{y}(t_k) - \bar{C}\bar{\chi}_d(t_k) ,
        \\
        \bar{\chi}_d(t_k) &= Q_{\Lambda_k}(\chi_d(t_k)), 
        \\
         u(t_k) \!\!\!\!\!\!&= \bar{K} \bar{\chi}_d(t_k),
     \end{cases} \label{a10}
\end{align}
where $\bar{A}_d = Q_{\Lambda^2}(A_d), \bar{B}_d = Q_{\Lambda}(B_d),\bar{L}_d = Q_{\Lambda}(L_d), \bar{K} = Q_{\Lambda}(K), \bar{C} = Q_{\Lambda}(C)$, and $\bar{y}(t_k) = Q_{\Lambda \Lambda_k}(y(t_k))$.

It is stressed that (\ref{a10}) is in the discrete form. To analyze the stability of the encrypted observer-based control for the continuous-time system, consider the dynamics of $\chi_v(t)$ as
\begin{align}
\begin{cases}
    \dot{\chi}_v(t) \!\!\!\!\!\!\!\!\!\!\!\!\!\! &= A_v \chi_v(t) \!+\! B_v u(t_k) 
    \\
     &~~~~~~~~~  + L_v (\bar{y}(t_k) \!-\! \bar{C} \bar{\chi}_v(t_k)  )  \!+\! {D} {\Tilde{\chi}_v(t_k)},
    \\
    \bar{\chi}_v(t_k) \!\!\!\!\! &= Q_{\Lambda_k}\left( \chi_v(t_k) \right),
    \\
     u(t_k) \!\!\!\!\!\!\!\!\!\! &= \bar{K} \bar{\chi}_v(t_k), \quad \forall t\in [t_k,t_{k+1}), \label{a11} 
\end{cases}
\end{align}
where, $\td{\chi}_v(t_k) = \bar{\chi}_v(t_k) - \chi_v(t_k)$, $A_v, B_v, L_v$  are defined as 
$e^{A_v h} = \bar{A}_d, \int_0^h e^{A_v \tau}d\tau B_v = \bar{B}_d, \int_0^h e^{A_v\tau}d\tau L_v = \bar{L}_d$, and $D = \left( \int_0^h e^{A_v\tau}d\tau \right)^{-1}\bar{A}_d$. 
To see the equivalence between \eqref{a11} and \eqref{a10} at each sampling instant, with $h=t_{k+1} - t_k$, we consider the solution of \eqref{a11} in the interval $[t_k, t_{k+1}]$ as
\begin{align*}
    \chi_v(t_{k+1}) &= \Bar{A}_d \chi_v(t_k) + \Bar{B}_d u(t_k) + \Bar{L}_d (\Bar{y}(t_k) - \Bar{C}\Bar{\chi}_v(t_k)) \non
    \\
    &~~~~~~~~~~~~~~~~~~~~~~~~~ + \bigg(\int_{0}^h e^{A_v \tau} d\tau \bigg) D \Tilde{\chi}_v(t_k) \non 
    \\
    &= \Bar{A}_d \bar{\chi}_v(t_k) + \Bar{B}_d u(t_k) + \Bar{L}_d (\Bar{y}(t_k) - \Bar{C}\Bar{\chi}_v(t_k)),
\end{align*}
which is the same as the first equation of \eqref{a10}. 
As a result, $\chi_v(t_k) = \chi_d(t_k) ~ (\forall t_k \in \mathcal{I})$ if they have the same initial conditions.
The controller \eqref{a11} is called the continuous-time virtual dynamics of \eqref{a10} and is only utilized for stability analysis.

To obtain the linear sampled-data closed-loop system, let us rewrite the first equation of \eqref{a11} as 
\begin{align}
    \dot{\chi}_v(t) = A\chi_v(t) + &B u(t_k) \!+\! L(y(t_k) \!-\! C\chi_v(t_k)) \!+\! \vartheta(t), \label{dynamic_chi_v}
\end{align}
with 
$
\vartheta(t) = \Delta A \chi_v(t) + \Delta B \bar{K} \bar{\chi}_v(t_k) + L \Tilde{y}(t_k) + \Delta L \bar{y}(t_k) - LC \Tilde{\chi}_v(t_k) - L{\Tilde{C}} 
\bar{\chi}_v(t_k) \!-\! \Delta L \bar{C} \bar{\chi}_v(t_k) + D\Tilde{\chi}_v(t_k)
$, $\td{y}(t) = \bar y(t) - y(t)$, $\Tilde{C} = \Bar{C} - C$, $\Delta A = A_v - A, \Delta B = B_v - B, \Delta L = L_v - L$.

Defining $e(t) = x(t)-\chi_v(t)$, from \eqref{Obs_Ctrl_uncryp} and \eqref{dynamic_chi_v}, one gets
\begin{align*}
    \dot{e}(t) = A e(t) - LC e(t_k) - \vartheta(t).
\end{align*}
By letting $z(t) = [e^\top(t),\chi_v^\top(t)]^\top \in \Rbb^{2n}$, the following closed-loop system is given in the same form of (\ref{sd_sys:org}) as
\begin{align}
    \dot{z}(t) = (\mathcal{A} + \Delta \mathcal{A}) z(t) + (\mathcal{A}_c + \Delta \mathcal{A}_c) z(t_k) + \eta(t), \label{closed_loop_dynamic}
\end{align}
where 
\begin{align*}
&\Ac = \bbm A & {0}\\ {0} & A\ebm\!\!, \Ac_c=\bbm -L C & {0} \\ L C & BK\ebm\!\!,~
\Delta \mathcal{A} = \bbm {0} & -\Delta A \\ {0} & \Delta A\ebm,
\\
&\Delta \mathcal{A}_c = \bbm -\Delta L C & (L+\Delta L)\Tilde{C} - \Delta B (K+\Tilde{K})\\ \Delta L C & \Delta B (K+\Tilde{K}) + B \Tilde{K} -(L+\Delta L) \Tilde{C}\ebm,
\end{align*}
and 
\begin{align}
    \eta(t) &= \bbm \Mc_{11} & \Mc_{12} \\ \Mc_{21} & \Mc_{22} \ebm  \bbm \td{\chi}_v(t_k) \\ \td{y}(t_k) \ebm  
     = \Mc \bbm \td{\chi}_v(t_k) \\ \td{y}(t_k) \ebm, \label{eta}
    \\
    \Mc_{11} &= -\Delta B(K+\Tilde{K})+(L+\Delta L) (C+\td{C})-D, \non
    \\
    \Mc_{21} &= (B + \Delta B)(K+\Tilde{K})-(L+\Delta L)(C+\Tilde{C})+D, \non
    \\
    \Mc_{12} &= -(L+\Delta L),  \Mc_{22} = L+\Delta L, \td K = \bar K - K.  \non
\end{align}
In the closed-loop system \eqref{closed_loop_dynamic}, $\Delta \Ac, \Delta \Ac_c$ are considered as the uncertainties, and $\eta(t)$ is considered as the disturbance of the system.
Obviously, the uncertainties and disturbance are directly associated with the quantizers, and their bounds can be chosen arbitrarily small by selecting large enough quantization gains.
Furthermore, the uncertainties $\Delta \Ac$ and $\Delta \Ac_c$ can be adjusted by choosing the appropriate static quantizers, while the disturbance $\eta(t)$ depends on the dynamic quantizer.

The following lemma provides a 
useful evaluation for bound of uncertainties in (\ref{closed_loop_dynamic}).
\begin{Lemma} \label{lemma_bound_of_Delta_ABL}
Let $\Tilde{A}_d = \bar{A}_d - A_d, \Tilde{B}_d = \bar{B}_d - B_d, \Tilde{L}_d = \bar{L}_d - L_d$, and assume that  $\|\Tilde{A}_d\| \le \gamma_A, \| \Tilde{B}_d \| \le \gamma_B$, $\| \Tilde{L}_d \| \le \gamma_L$, 
and small enough $h$ such that $\| e^{-Ah} \|\gamma_A < 1$.
Then, the following inequalities hold
\begin{subequations}
\begin{align}
\| \Delta A \| &\le  \frac{\|e^{-Ah}\|\gamma_A}{h(1-\|e^{-Ah}\|\gamma_A)} = \delta_A(h,\gamma_A), \\ 
\| \Delta B \| &\le \frac{\alpha}{1-\alpha \beta}\left( \gamma_B \!+\! \beta \|B\| \right) = \delta_B(h,\gamma_B,\gamma_A), \\
\| \Delta L \| &\le \frac{\alpha}{1-\alpha \beta} \left( \gamma_L \!+\! \beta \|L\| \right) = \delta_L(h,\gamma_L,\gamma_A),
\end{align} \label{bound_of_Delta_ABL}
\end{subequations}
where $\alpha = \left \|\left( \int_0^h e^{A\tau}d\tau \right)^{-1} \right\| > 0$, $\beta = \frac{e^{\|A\|h}h^2\delta_A}{1-\delta_A} > 0$.
\end{Lemma}
\textit{Proof:} See Appendix A.

\begin{Remark}
This paper exploits a particular case of Lemma 3, where
in accordance with the property \eqref{quantization_error_bound}, one has $\|\td{A}_d\| \le \frac{n}{2\Lambda^2}$, $\|\td{B}_d\| \le \frac{\sqrt{mn}}{2\Lambda}$, $\| \td{L}_d \| \le \frac{\sqrt{nr}}{2\Lambda}$. 
Then, by choosing $\gamma_A = \frac{n}{2\Lambda^2}$, $\gamma_B = \frac{\sqrt{mn}}{2\Lambda}, \gamma_L = \frac{\sqrt{nr}}{2\Lambda}$, we can assess the bounds of $\|\Delta A\|$, $\|\Delta B\|$ and $\|\Delta L\|$ as functions of $h$ and $\Lambda$, i.e., $\delta_A(h,\Lambda)$, $\delta_B(h,\Lambda)$ and $\delta_L(h,\Lambda)$.
\end{Remark}

\section{Stability Analysis} \label{stability_analys}
This section provides the stability analysis of the sampled-data system \eqref{closed_loop_dynamic}. In the subsection \ref{stability_part_A}, some results are presented for stability analysis of the system \eqref{sd_sys:org}. The only difference between the systems \eqref{sd_sys:org} and \eqref{closed_loop_dynamic} is the presence of the uncertainties in \eqref{closed_loop_dynamic}. In the subsection \ref{stability_part_B}, the LMIs-based conditions are proposed to cope with the uncertainties, then the results in the subsection \ref{stability_part_A} could be implemented.


\subsection{Modified discontinuous Lyapunov functional for sampled-data systems} \label{stability_part_A}
The following theorems provide sufficient conditions for the stability analysis of the sampled-data system \eqref{sd_sys:org}.
\begin{Theorem} \label{theorem_stability_sampled_data_sys}
Let $V(z): \Rbb^n \rightarrow \Rbb_+$ be a continuous differentiable function, and
there exist $\mu_1, \mu_2 >0$ such that $\mu_1 \| z \|^2 \le V(z) \le \mu_2 \| z \|^2 $.
Suppose that there exist differentiable functionals $\Uc(t,z)$, $\Wc(t,z)$ over $t\in \Rbb\setminus\Ic$, and positive scalars $\mu_3, \mu_4$  such that 
\begin{align}
     &\Uc^-_{k+1} - \Wc_k^+ \ge 0  \label{condi_1_for_functionals},
     \\
     &\dot{\Fc}(t,z)  \leq -\mu_3 \| z \|^2 + \mu_4 \| \eta \|^2 \label{condi_2_for_functionals},
\end{align}
where 
$\Fc(t,z) = V(z) + (t-t_k)\Uc(t,z) + (t_{k+1} - t)\Wc(t,z)$,
 $\Uc^-_k = \lim_{t\rightarrow t_k^-} \Uc(t,z(t))$ and $\Wc_k^+ = \lim_{t\rightarrow t_k^+}\Wc(t,z(t))$. 
Then, for the bounded energy disturbance $\eta(t)$
$\left( \int_0^{\infty} \big\| \eta(t) \big\|^2 dt < \Ec_\eta \right)$,
the system \eqref{sd_sys:org} satisfies IQC.
\end{Theorem}

\textit{Proof:} To begin with, let us take time derivative of $\Fc(t,z)$:
$\dot{\Fc}(t,z) = \dot{V}(z) + \Uc(t,z) - \Wc(t,z)  + (t-t_k)\dot{\Uc}(t,z)  + (t_{k+1} - t)\dot{\Wc}(t,z)$.
For simple notations, let $V_k = V(t_k,z(t_k))$. By integrating from $t_k$ to $t_{k+1}$ both sides of (\ref{condi_2_for_functionals}) along the solution of \eqref{sd_sys:org}, it yields
\begin{align*}
    \int_{t_k}^{t_{k+1}} \dot{\Fc}(t,z)dt &= \lim_{t\rightarrow t_{k+1}^{-}}\Fc(t,z(t)) - \lim_{t\rightarrow t_k^+} \Fc(t,z(t))
    \\
    &= V_{k+1} - V_k + h(\Uc^-_{k+1} - \Wc_k^+)
    \\
    &\leq -\mu_3 \int_{t_k}^{t_{k+1}} \!\!\!\!\! \| z(\tau) \|^2 d\tau 
    + \mu_4 \int_{t_k}^{t_{k+1}} \!\!\!\!\!      \| \eta(\tau) \|^2 d\tau.
\end{align*}
According to (\ref{condi_1_for_functionals}), we further obtain
\begin{align}
V_{k+1} - V_k + \mu_3 \int_{t_k}^{t_{k+1}} \!\!\!\!\! \| z(\tau) \|^2 d\tau
\leq \mu_4 \int_{t_k}^{t_{k+1}} \!\!\!\!\!      \| \eta(\tau) \|^2 d\tau.   \label{thm1:pf1} 
\end{align}
Summing up (\ref{thm1:pf1}) from $0$ to $k+1$, we have $\int_{0}^{t_{k+1}} \| z(\tau) \|^2 d\tau \leq \mu_3^{-1} \left( V_0 + \mu_4 \Ec_\eta \right)$. As a result, the system \eqref{sd_sys:org} satisfies the IQC defined in Definition \ref{Def1}.
$\Qced$

In Theorem \ref{theorem_stability_sampled_data_sys}, the disturbance $\eta(t)$ satisfies the bounded energy condition; it also means that $\eta(t) \rightarrow 0$ as $t\rightarrow \infty$. However, in some situations, the disturbance always exists and does not vanish. The following theorem considers the case that the disturbance $\eta(t)$ is only bounded, i.e., $\Vert \eta(t) \Vert \le \bar\eta,~ \forall t \ge 0 $ for some $\bar \eta >0$.
\begin{Theorem} \label{theorem2}
Suppose that the conditions (\ref{condi_1_for_functionals}) and (\ref{condi_2_for_functionals}) are satisfied, and the disturbance $\eta(t)$ is bounded by $\bar\eta >0$ as $\Vert \eta(t) \Vert \leq \bar\eta,~ \forall t \ge 0 $.
Then, if $z(t_k) \notin \Omega_\rho$ where
\begin{align}
\Omega_\rho &= \left\{ z \in\Rbb^n :  V(z) \leq \overline{V}_\rho \right\},~
\overline{V}_\rho  = \max_{z \in \Bc_\rho} V(z), \label{Omega_rs}
\\
\Bc_\rho &= \left\{ z \in\Rbb^n : \Vert z \Vert^2 \leq \rho  =  \frac{\mu_4 \bar\eta^2 + \sigma}{\mu_3}, 0 < \sigma  \right\},  \label{Bc_r}
\end{align}
the solution of (\ref{sd_sys:org}) in $[t_k, t_{k+q}]$ enters $\Omega_\rho$ at least once when $t_{k+q} \geq \frac{V_k - \overline{V}_\rho}{ \sigma} + t_k$.
\end{Theorem}
\textit{Proof:}  For $z_k \notin \Omega_\rho$, let $q \in \Nbb$ such that $z(t) \notin \Omega_\rho$ for all $t \in \Tc_{k,q} \triangleq [t_k, t_{k+q}]$.
Accordingly, $V(z(t)) \geq \ovl V_\rho =  \max_{z \in \Bc_\rho} V(z)$, and then 
$z(t) \notin \Bc_\rho$ for all $t \in \Tc_{k,q}$.
As a result, $\mu_3\Vert z(t)\Vert^2 - \mu_4 \bar\eta > \sigma$.
From (\ref{thm1:pf1}), we have that $V_{k+1} - V_k \leq (t_k - t_{k+1}) \sigma$.
The summation of the inequality from $t_k$ to $t_{k+q}$ results in
$V_{k+q} - V_k = \sum_{i = 0}^{q-1} (V_{k+i+1} - V_{k+i}) < (t_k - t_{k+q}) \sigma  $.
Thus, since $V_{k+q} > \ovl V_\rho$ with $t\in\Tc_{k,q}$, we have that
$V_k - \overline{V}_\rho > V_k - V_{k+q} > (t_{k+q} - t_k) \sigma$,
that is, $t_{k+q} - t_k < \frac{V_k - \overline{V}_\rho}{\sigma}$.
Therefore, for $z_k \notin \Omega_\rho$ and $t_{k+q} \geq t_k + \frac{V_k - \overline{V}_\rho}{\mu_3\sigma}$, $\exists t \in [t_k, t_{k+q}]$ such that $z(t) \in \Omega_\rho$.
$\Qced$

\subsection{Stability analysis of the closed-loop system} \label{stability_part_B}
This subsection provides sufficient conditions for the stability of the system \eqref{closed_loop_dynamic}. 
For convenience, let us denote
\begin{align*}
    &\xi(t) = \bbm z^\top(t) & z^\top(t_k) & \phi^\top_k(t)\ebm^\top \in \Rbb^{6n},
    \\
    &\phi_k(t) = \frac{1}{t-t_k}\int_{t_k}^t z(\tau) d\tau \!\in\! \Rbb^{2n}, 
    \Phi_0 \!=\! [\Ac, \Ac_c, {0}] \in \Rbb^{2n\times 6n}, 
    \\
    &E_1 = \! \bbm I \!& \!0\! & \!0\! \ebm, ~ E_2 \!=\! \bbm 0\!&\!I\!&\!0 \ebm, E_3 \!=\! \bbm 0\!&\!0\!&\!I\! \ebm \in \Rbb^{2n\times 6n},
    \\
    &\Phi_1 = E_1 - E_2, \Phi_2 = E_1 + E_2 -2E_3,
    \\
    &\Phi_3  = \bbm{0} & \Ac_c & \Ac\ebm \in \Rbb^{2n \times 6n}, \Phi_4 = [E_1^\top ~~ E_2^\top]^\top,
    \\
    &\Phi_5 = \bbm \Phi_0^\top & {0} \ebm^\top \in \Rbb^{4n \times 6n},
    \Phi_6  = \bbm I & 0 \ebm^\top \in \Rbb^{4n\times 2n},
    \\
    &\Delta_0 = \bbm\Delta \Ac & \Delta \Ac_c & {0}\ebm,
    \Delta_2  = \left[ \Delta_5^\top~ \Delta_0^\top~ \Delta_3^\top \right]^\top\!\!,
    \\
    &\Delta_3 = [{0},\Delta \Ac_c, \Delta \Ac] \in \Rbb^{2n \times 6n},
    \Delta_5  = \bbm\Delta_0^\top & {0}\ebm^\top \in \Rbb^{4n \times 6n}.
\end{align*}
Considering the uncertainty terms in \eqref{closed_loop_dynamic}, it follows that 
\begin{align}
    \|\Delta \Ac\| &= \sqrt{2}\|\Delta A\| \le \sqrt{2}\delta_A(h,\Lambda), \label{bound_of_Delta_Ac}
    \\
    \|\Delta \Ac_c\| &\le 2\| \Delta L \|\|C\| + (\|L\| + \|\Delta L\|)\|\td{C}\| \nonumber
    \\
    &\quad+ 2\|\Delta B\|(\|K\|+\|\td{K}\|) + \|B\| \|\td{K}\| \nonumber
    \\
    &\quad+ \|\td{C}\| (\|L\| + \|\Delta L\|) \nonumber
    \\
    &\le 2\left( \|C\| + \frac{\sqrt{nr}}{2\Lambda} \right) \delta_L(h,\Lambda) \nonumber
    \\
    &\quad+ \frac{\sqrt{nr}\|L\|}{2\Lambda} + 2\left( \|K\| + \frac{\sqrt{mn}}{2\Lambda}\right)\delta_B(h,\Lambda) \nonumber
    \\
    &\quad+ \frac{\sqrt{mn}\|B\|}{2\Lambda} = \varphi(h,\Lambda). \label{bound_of_Lambda_Ac_c}
\end{align}
Let a positive value $\Mc_U$ be an upper bound of $\|\Mc\|$; then, from \eqref{eta}, we have
\begin{align}
    \| \eta(t) \| &\le \Mc_U \big( \|\Tilde{\chi}_v(t_k)\| + \|\Tilde{y}(t_k)\|\big) \nonumber \\
    &\le \Mc_U \left( \frac{\sqrt{n}}{2} + \frac{\sqrt{r}}{2\Lambda} \right) \frac{1}{\Lambda_k}. \label{b22}
\end{align}
Thus,
\begin{align}
    \int_0^{\infty} \|\eta(t)\|^2 dt &\le \int_0^{\infty} \Mc_U^2  \left( \frac{\sqrt{n}}{2} + \frac{\sqrt{r}}{2\Lambda} \right)^2 \frac{1}{\Lambda_k^2} dt \nonumber 
    \\
    &= h\Mc_U^2  \left( \frac{\sqrt{n}}{2} + \frac{\sqrt{r}}{2\Lambda} \right)^2 \sum_{k=0}^{\infty} \frac{1}{\Lambda_k^2}. \label{bound_of_eta}
\end{align}
The following theorem provides sufficient conditions for the stability of the system \eqref{closed_loop_dynamic} based on LMIs.
\begin{Theorem} \label{thm3}
Assume that there exist $h, \varepsilon_1, \varepsilon_2, \epsilon_1, \epsilon_2 \in \Rbb_{++}$, matrices $U_1 \in \Sbb^{4n\times 4n}$; $F, W_1, H, U_4 \in \Sbb^{2n \times 2n}$; $P,R \in \Sbb^{2n\times 2n}_{++}$; $U_2, U_3, W_2 \in \Rbb^{2n\times 2n}$, $\Psi \in \Sbb^{6n\times6n}_{++}$, and $Q, N_1, N_2 \in \Rbb^{2n\times6n}$ such that 
\begin{align}
    0 &\preceq \Ubf - E_2^\top H E_2, \label{condi_1_main_theorem}
    \\
    0 &\succeq \bbm  
    \Xi_0 + h \Xi_1 + \varepsilon_1 I & Y_{01} \\
    Y_{01}^\top  & -\epsilon_1 I
    \ebm, \label{condi_2_main_theorem}
    \\
    0 &\succeq h R - \epsilon_1 I, \label{condi_3_main_theorem}
    \\
    0 &\succeq \bbm
    \Xi_0 + h\Xi_2 + \varepsilon_2 I & (\ast) & (\ast) & (\ast) \\
    h N_1 & -h R & (\ast) & (\ast) \\ 
    h N_2 & {0} & -3h R & (\ast) \\
    Y_{02}^\top & {0} & {0} & -\epsilon_2 I
    \ebm, \label{condi_4_main_theorem}
\end{align}
where 
\begin{align*}
    \Ubf =& \bbm 
    U_1 & (\ast) \\
    \bbm U_2 & U_3 \ebm & U_4
    \ebm,
    \\
    \Xi_0 =&~ {\bf He}\Big\{E_1^\top P \Phi_0 \Big\} \!+\! U \!-\! \Phi_1^\top W_1 \Phi_1 \!-\! {\bf He} \Big\{ \Phi_1^\top W_2 E_2 \Big\} \\
    &-E_1^\top H E_1 \!+\! {\bf He} \Big\{ \!-\! \Phi_1^\top N_1 \!-\! \Phi_2^\top N_2 \!+\! (E_1-E_3)^\top \\
    &\times (U_2 E_1 + U_3 E_2 + U_4 E_3) + \Phi_1^\top Q \Big\} + \Psi,
    \\
    \Xi_1 =&~ {\bf He} \Big\{\Phi_0^\top(W_1(E_1-E_2) + W_2E_2 + HE_1)\Big\} \\
    &+ E_2^\top F E_2 + \Phi_0^\top R \Phi_0,
    \\
    \Xi_2=& - E_2^\top F E_2 + {\bf He} \Big\{\Phi_4^\top U_1 \Phi_5 + E_3^\top U_2 \Phi_0 - \Phi_3^\top Q \Big\},
    \\
    Y_0 =&~ E_1^\top P,~ Y_1 = (E_1 \!-\! E_2)^\top W_1 \!+\! \Phi_0^\top R,~ Y_{01} = Y_0 \!+\! h Y_1,
    \\
    Y_2 =& \bbm \Phi_4^\top U_1 & E_3^\top U_2 & -Q^\top \ebm, Y_{02} = \bbm Y_0 & hY_2 \ebm;
\end{align*}
and the value $\Lambda$ is chosen such that 
\begin{align}
    2\delta_A^2(h,\Lambda) + \varphi^2(h,\Lambda) \le \min \Big\{ \frac{\varepsilon_1}{2\epsilon_1}, \frac{\varepsilon_2}{6 \epsilon_2} \Big\}.   \label{condition_for_Lambda}
\end{align}
\begin{itemize}
    \item[1)] If the value $\Lambda_k$ is chosen such that 
\begin{align}
    \sum_{k=0}^\infty \frac{1}{\Lambda_k^2} < \infty, \label{condition_for_Lambda_k}
\end{align}
then, the system \eqref{closed_loop_dynamic} is globally asymptotically stable. \label{theorem3_1}
\item[2)] If $\Lambda_k$ is a fixed value, there exist $\mu_3, \mu_4, \sigma > 0$, $q \in \Nbb$ such that any solution of \eqref{closed_loop_dynamic} in $[t_k, t_{k+q})$ enters $\Omega_{\rho}$ at least once where $\Omega_{\rho}$ is defined in \eqref{Omega_rs} and \eqref{Bc_r}, and $\bar \eta$ is obtained from \eqref{b22}. \label{theorem3_2}
\end{itemize}
\end{Theorem}
\textit{Proof:} 
See Appendix B.
\begin{Remark}
    By preselecting the sampling interval $h$, the conditions \eqref{condi_1_main_theorem} - \eqref{condi_4_main_theorem} can be considered as the LMIs. 
    Thus, Theorem \ref{theorem3_1} provides a stability criterion for the selection of sampling interval $h$. 
    Moreover, through the inequality (\ref{condition_for_Lambda}), Theorem \ref{theorem3_1} exposes a relationship between $h$ and $\Lambda$ to ensure the stability of the encrypted control system.
    By checking the feasibility of the LMIs in \eqref{condi_1_main_theorem}-\eqref{condi_4_main_theorem}, we can obtain the range for $h$ such that there exists at least one quantization scheme such that the system \eqref{closed_loop_dynamic} is stable.
\end{Remark}

\section{Evaluation} \label{simulation}
This section considers the angular control problem for DC motor with the dynamics described as (see \cite{DCmotor})
\begin{subequations}
    \begin{align}
        \frac{d i_a (t)}{dt} &= -\frac{R_a}{L_a} i_a (t) - \frac{k_d}{L_a} \omega(t) + 
        \frac{1}{L_a} v_s(t), \label{DC_motor_eq1}
        \\
        \frac{d \omega(t)}{dt} &= \frac{k_d}{J_M} i_a(t) - \frac{B_M}{J_M} \omega(t), \label{DC_motor_eq2}
        \\
        \frac{d \theta(t)}{dt} &= \omega(t), \label{DC_motor_eq3}
    \end{align} \label{DC_motor}
\end{subequations}
where $i_a, v_s, \theta$ and $\omega$ represent the armature current, armature voltage, angular position, and angular velocity of the rotor, respectively. The system's parameters include the armature resistance $R_a = 7.2 [\Omega]$, the armature inductance $L_a = 0.0917[H]$, the frictional constant $B_M = 0.0004 [N.m.s/rad]$, the torque constant $k_d = 0.1236 [N.m/Wb.A]$, and the moment of inertia $J = 0.0007046 [kg.m^2]$.

Our target is to derive the DC motor to a reference angle, i.e., $\theta(t) \rightarrow \theta_r$, where $\theta_r$ is the desired angle and is assumed to be a constant. 
Define the angular position error as $\theta_e(t) = \theta(t) - \theta_r$, the equation \eqref{DC_motor_eq3} can be rewritten as $\dot{\theta}_e(t) = \omega(t)$. 
Denoting the state variable $x(t) = \bbm i_a(t) & \omega(t) & \theta_e(t)\ebm^\top$ and the output $y(t) = \theta_e(t)$, the system dynamic \eqref{DC_motor} can be rewritten in the state space form as 
\begin{subequations}
    \begin{align}
        \dot{x}(t) &= A x(t) + B u(t),
        \\
        y(t) &= C x(t),
    \end{align} \label{DC_LTI}
\end{subequations}
where 
\begin{align*}
    A = \bbm 
         -\frac{R_a}{L_a} & -\frac{k_d}{L_a} & 0
         \\
         \frac{k_d}{J_M} & -\frac{B_M}{J_M} & 0
         \\
         0 & 1 & 0
        \ebm, 
   B = \bbm \frac{1}{L_a} \\ 0 \\ 0 \ebm, 
   C = \bbm 0 & 0 & 1 \ebm.
\end{align*}
By pole-placement method \cite{antsaklis2007linear}, let us choose $K = [ 1.65, -6.26, -43.08 ]$, and $L = [ 69.11, 71.91, 24.13 ]^\top$.

With such setups, Theorem \ref{thm3} can provide feasible solutions with maximal sampling interval  $\overline{h}$ up to $0.083$. 
In addition,
the minimal values for $\Lambda$ corresponding to several sampling intervals are given in Table \ref{Table1}.
It can be seen from Table \ref{Table1} that the larger the sampling period is, the larger value of $\Lambda$ is needed.
That is, the quantization is required to be more accurate when the sampling interval becomes larger.
\begin{table}[]
\caption{ Minimal values of $\Lambda$ provided by \eqref{condition_for_Lambda} corresponding to several choices of sampling interval $h$.}
\centering
\begin{adjustbox}{width= 0.48\textwidth,center}
\begin{tabular}{l|  l  l  l  l  l }
\hline \hline
$h$  & $0.083$                      & $0.07$                       & $0.05$                       & $0.03$                       & $0.01$                    \\
\hline
$\Lambda$ & $1.92 \times 10^5$ & $2.15\times 10^4$ & $9.88 \times 10^3$ & $6.01 \times 10^3$ & $4.16 \times 10^3$
\\
\hline \hline 
\end{tabular} 
\end{adjustbox}
\label{Table1}
\end{table}
\begin{table}[]
\caption{ MRMS values corresponding to several choices of varying values of $\Lambda_k$ at $t = 53[s]$ with the window length $T_\text{RMS} =50[s]$.}
\centering
\begin{adjustbox}{width= 0.49\textwidth,center}
\begin{tabular}{l|  l  l  l  l  l}
\hline \hline
$\Lambda_k$  & $k^{0.4}$ & $k^{1}$ & $k^{1.5}$ & $k^2$ & $k^{3}$
\\
\hline
{MRMS} & $4.2 \!\times\! 10^{\!-\!2}$ & $1.7 \!\times\! 10^{\!-\!3}$ & $1.5 \!\times\! 10^{\!-\!4}$ & $3.2 \!\times\! 10^{\!-\!5}$ & $2.5 \!\times\! 10^{\!-\!7}$
\\
\hline \hline 
\end{tabular} 
\end{adjustbox}
\label{Table2}
\end{table}
\begin{table}[]
\caption{ MRMS values corresponding to several choices of fixed values of $\Lambda_k$ at $t = 53[s]$ with the window length $T_\text{RMS} =50[s]$.}
\centering
\begin{adjustbox}{width= 0.49\textwidth,center}
\begin{tabular}{l|  l  l  l  l  l}
\hline \hline
$\Lambda_k$  & $10$ & $30$ & $50$ & $70$ & $100$
\\
\hline
MRMS & $4.8\!\times\! 10^{\!-\!2}$ & $1.5 \!\times\! 10^{\!-\!2}$ & $9.7\!\times\! 10^{\!-\!3}$ & $6.2 \!\times\! 10^{\!-\!3}$ & $4.2\!\times\! 10^{\!-\!3}$
\\
\hline \hline 
\end{tabular} 
\end{adjustbox}
\label{Table3}
\end{table}

\begin{figure}
     \centering
     \begin{subfigure}[b]{\linewidth}
         \centering
         \includegraphics[width=\textwidth]{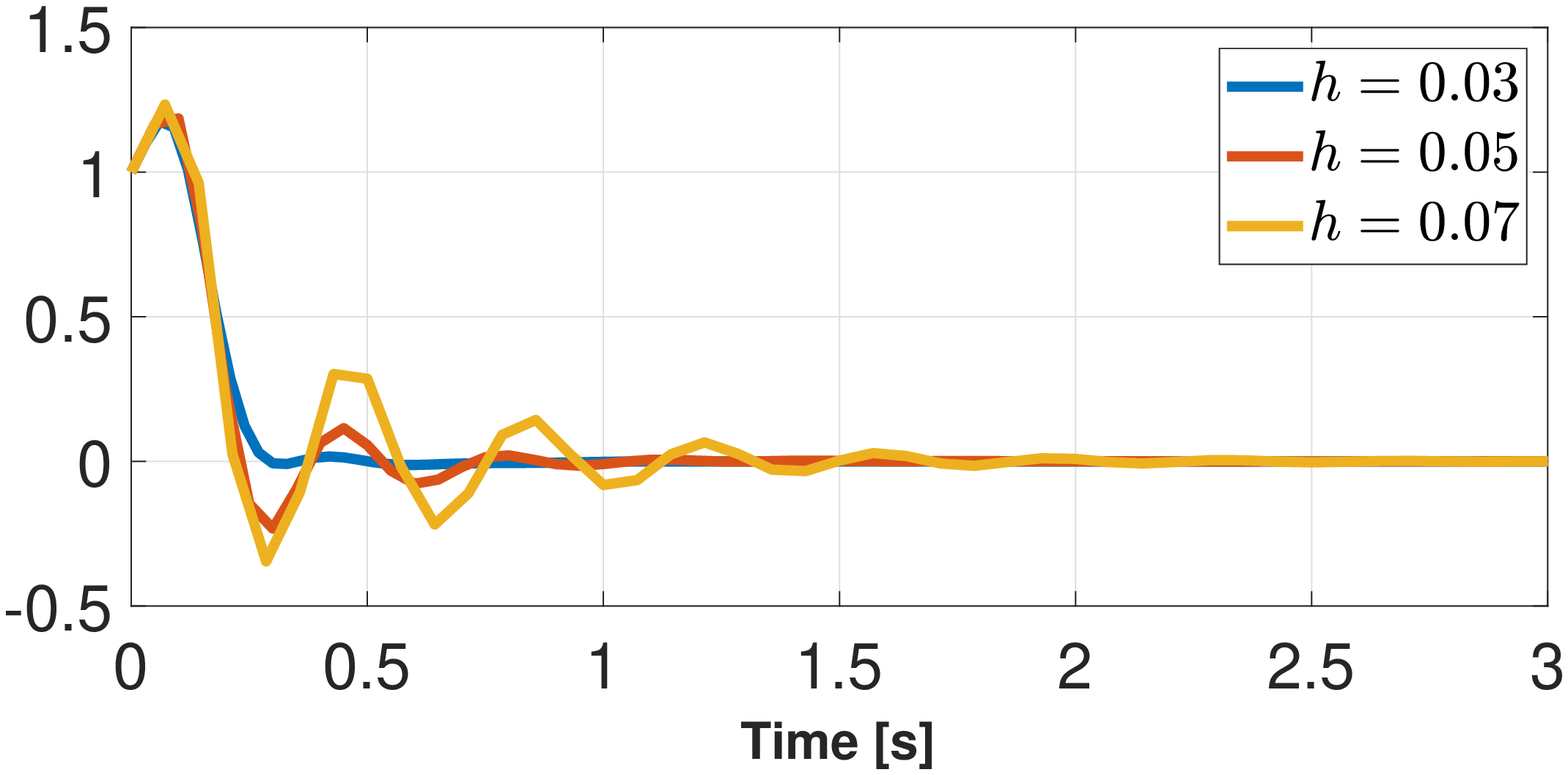}
         \caption{Angular position error $\theta_e [rad]$.}
         \label{theta_e_3cases}
     \end{subfigure}
     \begin{subfigure}[b]{\linewidth}
         \centering
         \includegraphics[width=\textwidth]{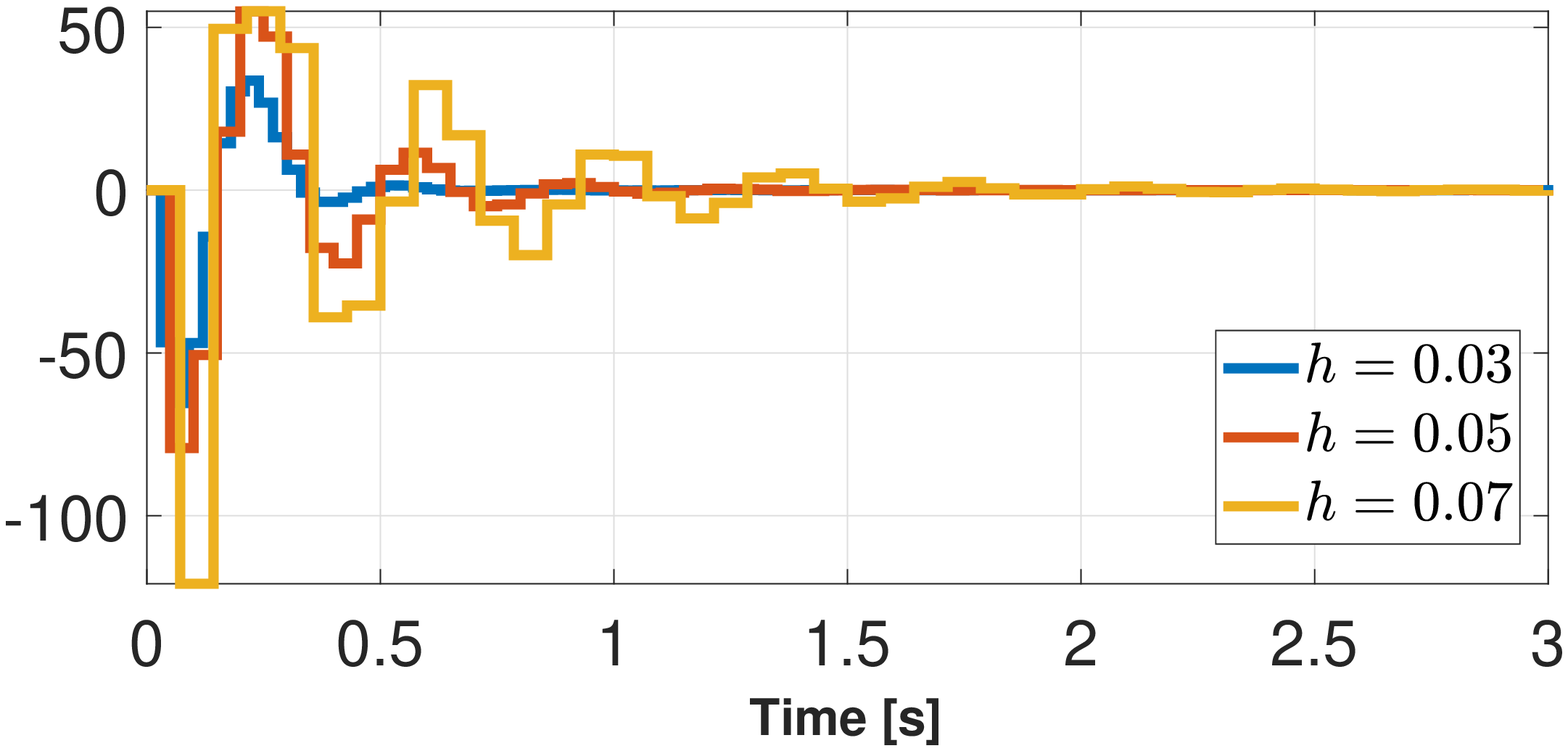}
         \caption{Control input $u [V]$.}
         \label{u_3cases}
     \end{subfigure}
        \caption{The angular position error and control input of the system in three cases $h = 0.03, 0.05, 0.07$ in accordance with $\Lambda$ shown in Table \ref{Table1}, and $\Lambda_k = k^2$.}
        \label{3cases}
\end{figure}
\begin{figure}
     \centering
     \begin{subfigure}[b]{0.241\textwidth}
         \centering
         \includegraphics[width=\textwidth]{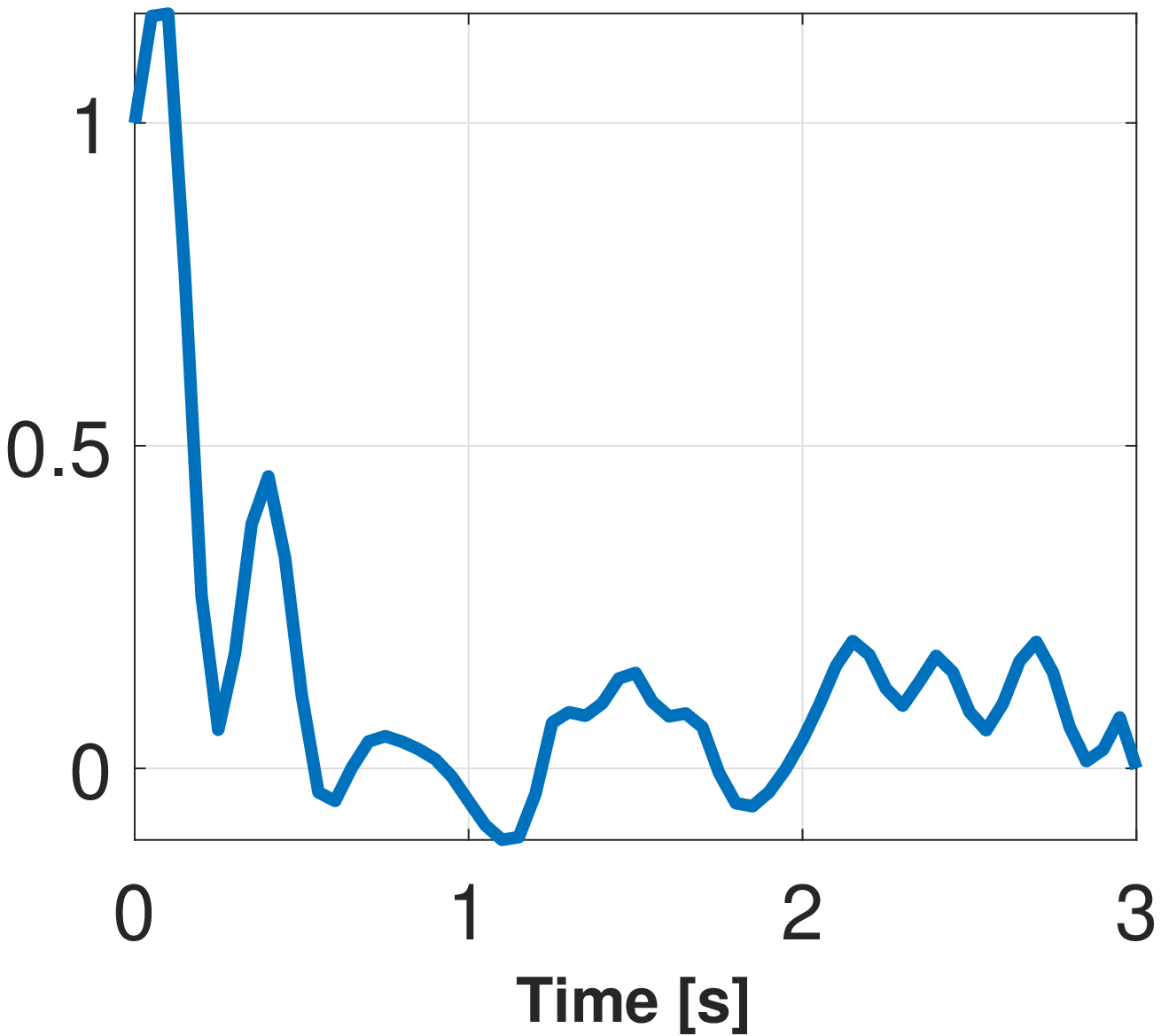}
         \caption{$\Lambda_k = k^{0.4}$.}
         \label{k0.4}
     \end{subfigure}
     \begin{subfigure}[b]{0.241\textwidth}
         \centering
         \includegraphics[width=\textwidth]{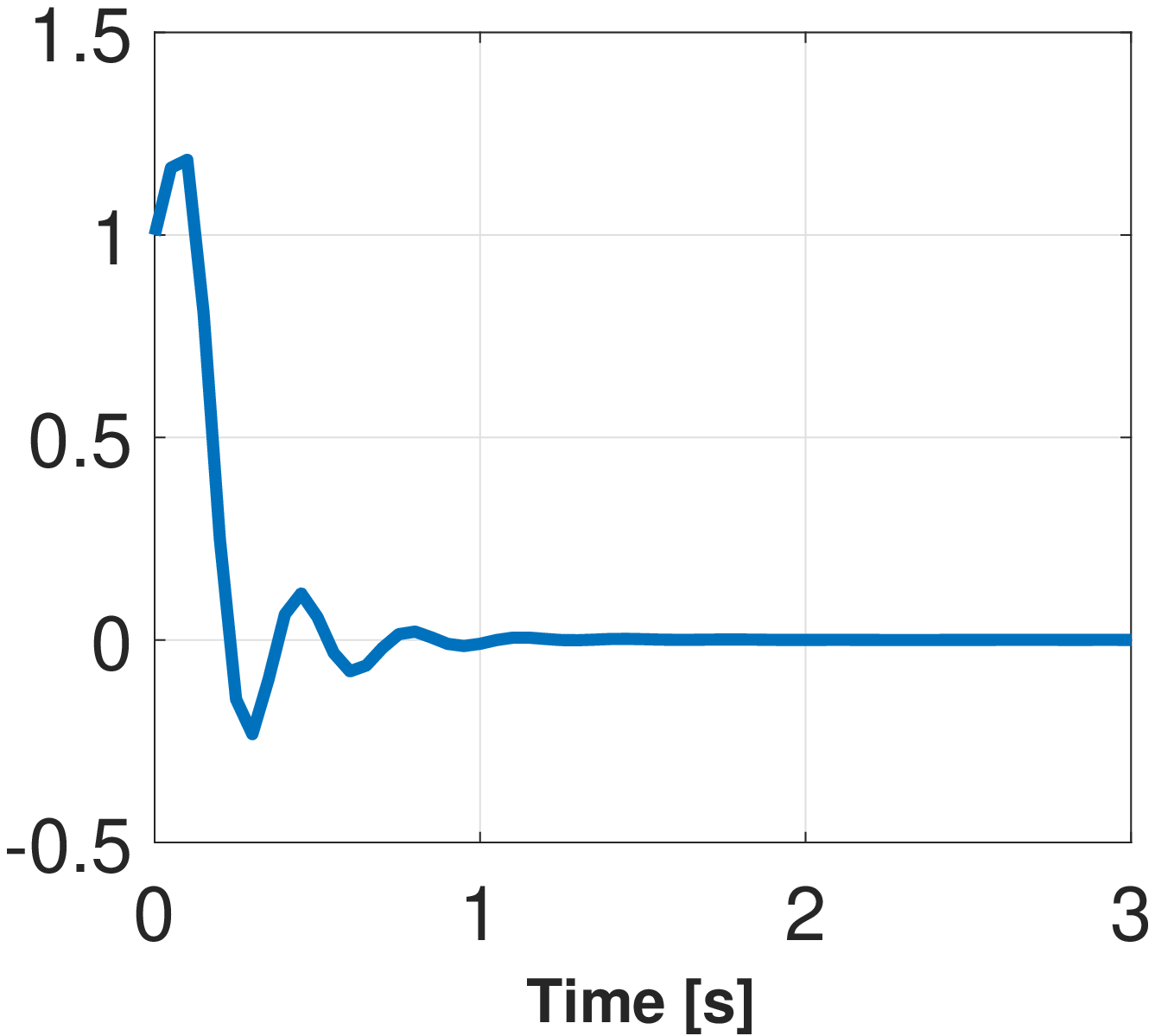}
         \caption{$\Lambda_k = k^{2}$.}
         \label{k2}
     \end{subfigure}   
        \caption{The angular position error $\theta_e [rad]$ with different time-varying $\Lambda_k$, $h = 0.05$, and $\Lambda = 9.88 \times 10^3$.}
        \label{varying_lambda_k}
\end{figure}
\begin{figure}
     \centering
     \begin{subfigure}[b]{0.241\textwidth}
         \centering
         \includegraphics[width=\linewidth]{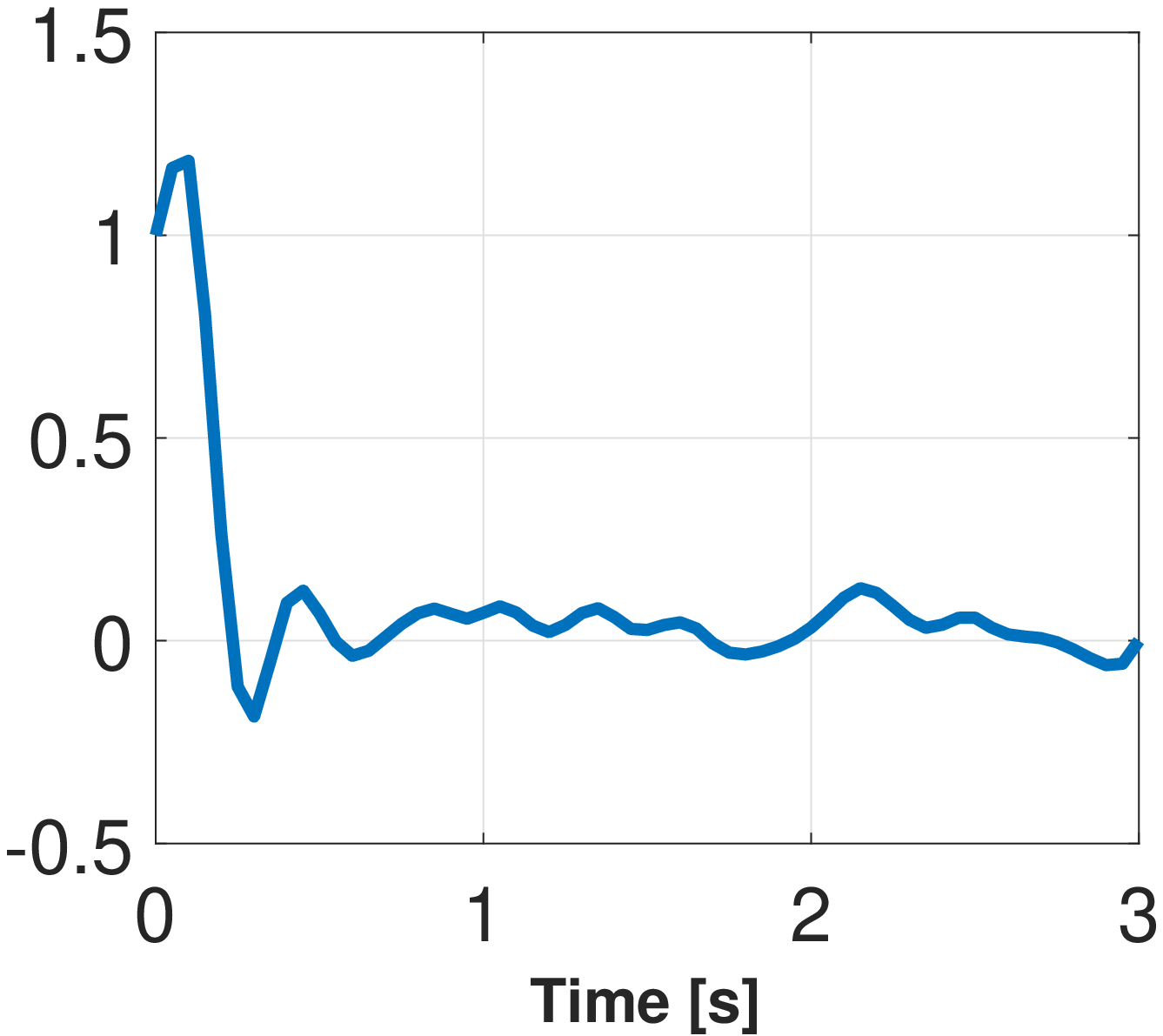}
         \caption{$\Lambda_k = 10$.}
         \label{k50}
     \end{subfigure}
     \begin{subfigure}[b]{0.241\textwidth}
         \centering
         \includegraphics[width=\linewidth]{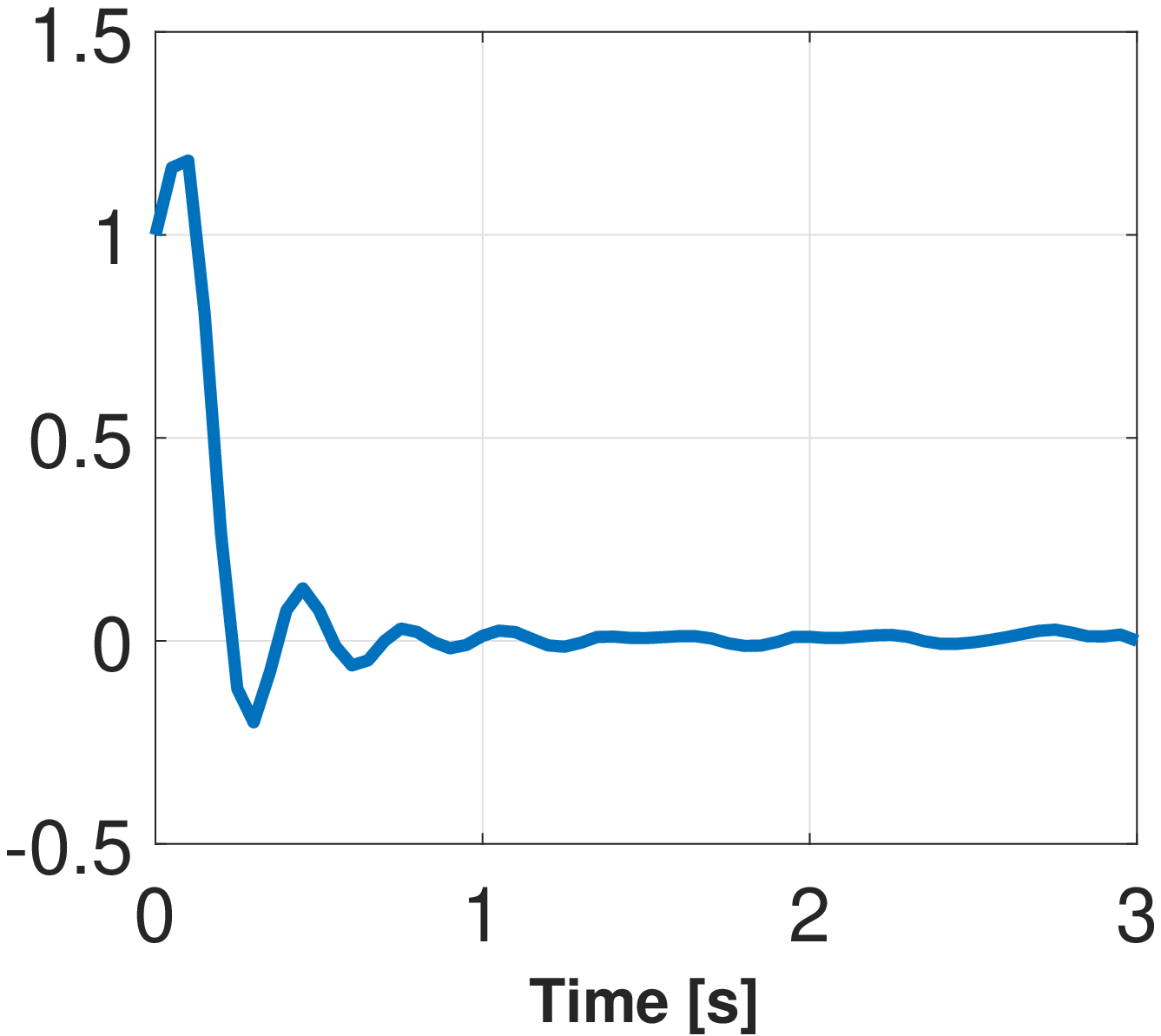}
         \caption{$\Lambda_k = 30$.}
         \label{k100}
     \end{subfigure}
        \caption{The angular position error $\theta_e [rad]$ with fixed values of $\Lambda_k$, $h = 0.05$, and $\Lambda = 9.88 \times 10^3$.}
        \label{fix_lambda_k}
\end{figure}
For more detail, let us consider three cases $h = 0.03, h = 0.05$, and $h=0.07$. The corresponding values for $\Lambda$ are chosen according to Table \ref{Table1} as $\Lambda = 6.01 \times 10^3, \Lambda = 9.88 \times 10^3$ and $\Lambda = 2.15 \times 10^4$; and the value $\Lambda_k$ for the dynamic quantizer is chosen as $\Lambda_k = k^2$. 
The results in Fig. \ref{3cases} figure out that choosing a smaller sampling period $h$ gives better results in terms of both control input and output. 
To be specific, the increment of $h$ leads to fluctuation in the angular position error and also increases the convergence time. Further, a larger value of $h$ needs a larger maximal value of the control input that could exceed the physical limitations of the control system.

Apart from this, Fig. \ref{varying_lambda_k} presents the time evolution of the angular position errors with different $\Lambda_k$.
The sampling period is chosen as $h=0.05$ with $\Lambda = 9.88 \times 10^3$.
We choose $\Lambda_k = k^{p}$ with $p > 0$; 
it is obvious that, if $p > 0.5$, then the condition \eqref{condition_for_Lambda_k} can be ensured. Fig. \ref{varying_lambda_k} confirms that larger $p$ results in better convergence of the angular position error.
Specifically, with $p = 0.4$, which does not satisfy the condition \eqref{condition_for_Lambda_k}, asymptotic convergence cannot be achieved (see Fig. \ref{k0.4}) while, with $p = 2$ which satisfy \eqref{condition_for_Lambda_k}, asymptotic convergence can be guaranteed as in Fig. \ref{k2}.
In Fig. \ref{fix_lambda_k}, we use the same setup as in the case of Fig. \ref{varying_lambda_k}, but $\Lambda_k$ is fixed to $10$ and $30$. 
As we can see, Fig. \ref{k100} with a larger value of $\Lambda_k$ ($\Lambda_k = 30$) shows better results compared to Fig. \ref{k50} with a smaller $\Lambda_k$ ($\Lambda_k=10$).

To see the asymptotic stability property of the system in accordance with $\Lambda_k$, let us introduce the moving root mean square (MRMS) \cite{kenney1939mathematics}, which is computed by the mean square of a signal over a given window length.
For a function $\theta_e(t)$, the MRMS with window length $T_\text{RMS}>0$ at time $t > T_\text{RMS}$ can be computed as
\begin{align}
   \text{MRMS} =  \sqrt{\frac{1}{T_\text{RMS}}\int_{t-T_\text{RMS}}^{t} \theta_e^2(\tau) d\tau }. \label{rms}
\end{align}
The MRMS computed in \eqref{rms} is used to measure the difference between $\theta_e(t)$ and $0$ during $[t-T_\text{RMS},t)$.
Let us consider the MRMS of $\theta_e(t)$ at $t=53[s]$ with the window length $T_\text{RMS} = 50[s]$.
Table \ref{Table2} shows the MRMS values according to $\Lambda_k = k^p$, with $p\in \{ 0.4,1.0,1.5,2.0,2.5 \}$. It can be seen that, with a larger $p$ which leads to larger values and a higher divergence speed of $\{\Lambda_k\}$, the MRMS is smaller. 
Accordingly, with a larger $p$, the angular position error $\theta_e(t)$ is closer to the horizontal axis which results in better convergence of the angular position error.
Similarly, with fixed values of $\Lambda_k$, Table \ref{Table3} shows that the larger $\Lambda_k$ is, the smaller MRMS can be computed; that is, a larger value of $\Lambda_k$ results in better convergence of the angular position error. 

\section{Conclusions} \label{conclusions}
This paper provided the stability analysis of continuous LTI systems with an encrypted observer-based controller. Due to encryption, computations for the observer-based controller were conducted based on its discrete-time model. Thus, the continuous-time virtual dynamics of the controller was also introduced for the stability analysis.
Due to the fact that the encryption-decryption process does not affect the stability of the system, we formulated the sampled-data closed-loop system with the presence of the uncertainties and disturbance associated with the static quantizers and the dynamic quantizer, respectively. 
Based on the discontinuous Lyapunov functional, we theoretically proved that, with suitable selections of the quantization gains and sampling interval, the system is globally asymptotically stable.
For future works, we will consider the same design for the aperiodic sampled-data system and provide more stability analysis in the case with only static quantizers.

 \section*{Acknowledgment}
This work was supported by the National Research Foundation
of Korea (NRF) grant funded by the Korea government (MSIT)
(2022R1A2B5B03001459).

\section*{Appendix}

\subsection{Proof of Lemma \ref{lemma_bound_of_Delta_ABL}} \label{proofLemma3}
First, $\Tilde{A}_d$ could be written as $\Tilde{A}_d = e^{A_v h}-e^{Ah} = e^{Ah}(e^{(A_v-A)h}-I)$; thus, $e^{{\Delta}A h}-I = e^{-Ah} \Tilde{A}_d$.
 One can obtain the error $\Delta A$ as: $\Delta A = \frac{1}{h} \ln({e^{-Ah} \Tilde{A}_d + I})$ with matrix logarithm defined in \cite{higham2008functions}. 
 Note that, for any matrix $M$ with $\| M-I \|<1$, we always have
\begin{align*}
    \ln{M} = \sum_{k=1}^{\infty}(-1)^{k+1} \frac{(M-I)^k}{k}.
\end{align*}
Then, with $\gamma_A$ and $h$ chosen such that $\| e^{-Ah} \|\gamma_A < 1$, we have
\begin{align*}
    \| \Delta A \| &= \frac{1}{h}\left\| \sum_{k=1}^{\infty} (-1)^{k+1} \frac{(e^{-Ah}\Tilde{A}_d)^k}{k}\right\| \le \frac{1}{h} \sum_{k=1}^{\infty} \left(\|e^{-Ah}\|\gamma_A\right)^k \\
    &= \frac{\|e^{-Ah}\|\gamma_A}{h(1-\|e^{-Ah}\|\gamma_A)} = \delta_A(h,\gamma_A).
\end{align*}
Next, we prove that $\| \Delta B \|, \| \Delta L \|$ are bounded by $\delta_B, \delta_L$. First, we have 
\begin{align*}
    \Tilde{B}_d = \bar{B}_d - B_d = \int_{0}^h e^{A_v \tau} d\tau B_v - \int_{0}^h e^{{A}\tau} d\tau  {B} 
    \\
    = \int_{0}^h \left( e^{A_v\tau} - e^{A\tau} \right)d\tau {B_v} + \int_0^h e^{A\tau} d\tau \Delta B.
\end{align*}
It can be seen that $\int_0^h e^{A\tau}d\tau$ is nonsingular, one gets
\begin{align*}
    \Delta B = \left( \int_0^h e^{A\tau}d\tau \right)^{-1} \! \left(\! \Tilde{B}_d - \int_0^h \left( e^{A_v\tau} - e^{A\tau} \right)d\tau  {B_v} \!\right).
\end{align*}
Therefore,
\begin{align}
    &\| \Delta B \| \le \alpha \left( \| \Tilde{B}_d \| + \left \| \int_0^h\left( e^{A_v\tau} - e^{A\tau} \right) d\tau\right \| \|B_v\|\right) \nonumber 
    \\
    &\le \alpha \left( \!\| \Tilde{B}_d \| \!+\! \int_0^h\!\left \| \left( e^{A_v\tau} \!-\! e^{A\tau} \right) \right \| d\tau (\|B\| + \|\Delta B\|)\!\right)\!. \label{b14}
\end{align}
Further, for all $\tau\in [0,h]$, we have
\begin{align*}
    \|e^{A_v\tau} &-e^{A\tau}\| = \| e^{A\tau} \left( e^{\Delta A\tau}\!-\!I \right) \| \le \left\|e^{A\tau}\right\| \! \left\| e^{\Delta A\tau} \!-\! I \right\|
    \\
    &\le e^{\|A\tau\|} \left\| \sum_{k=1}^{\infty} \frac{(\Delta A\tau)^k}{k} \right\| \le e^{\|A\|h} \sum_{k=1}^{\infty} (\|  \Delta A\|h)^k 
    \\
    &= e^{\|A\|h} \frac{\delta_A h}{1-\delta_A}.
\end{align*}
Substituting the above result into \eqref{b14} gives 
\begin{align*}
    \|\Delta B\| \le \frac{\alpha}{1-\alpha \beta} \left( \gamma_B + \beta \|B\| \right) = \delta_B(h,\gamma_B,\gamma_A).
\end{align*}
By a similar way, we also have $\|\Delta L\| \le \delta_L(h,\gamma_L,\gamma_A)$, which completes the proof.$\Qced$

\subsection{Proof of Theorem 3} \label{proofTheorem3}
Based on Theorem \ref{theorem_stability_sampled_data_sys}, select the following functions
\begin{subequations}
\begin{align}
    V(z) =&~ z^\top (t)Pz(t), \label{lyapunov_function_V}
    \\
    \Uc(t,z) =&~ \zeta^\top (t) U_1 \zeta(t) + 2 \phi_k^\top (t)(U_2 z(t) + U_3 z(t_k)) \nonumber \\
    &+ \phi_k^\top (t) U_4 \phi_k(t), \label{functional_Uc}
    \\
    \Wc(t,z) =&~ (t-t_k)z^\top(t_k) F z(t_k) + z^\top(t) H z(t) \nonumber \\
    &+\! \big( z(t) \!-\! z(t_k) \big)^\top \big( W_1(z(t) \!-\! z(t_k)) \!+\! 2W_2 z(t_k) \big) \nonumber \\
    &+\! \int_{t_k}^t \dot{z}^\top(\tau) R \dot{z}(\tau) d\tau, \label{functional_Wc}
\end{align}
\end{subequations}
where $\zeta(t) = \bbm z^\top(t) & z^\top(t_k) \ebm^\top$. It is obvious that the condition \eqref{condi_1_for_functionals} is satisfied under the condition \eqref{condi_1_main_theorem}. Taking the time derivatives of \eqref{lyapunov_function_V},\eqref{functional_Uc} and \eqref{functional_Wc} gives
\begin{subequations}
\begin{align}
    \dot{V} =&~  \xi^\top(t) {\bf He} \Big\{ E_1^\top P (\Phi_0 + \Delta_0) \Big\}\xi(t) \nonumber \\
     &+ 2\eta^\top(t) P E_1 \xi(t), \label{derivative_of_V}
     \\
    \dot{\Uc} =&~ \xi^\top(t){\bf He} \Big\{ \Phi_4^\top U_1 (\Phi_5+\Delta_5) + E_3^\top U_2 (\Phi_0 + \Delta_0) \nonumber \\
    &+ \frac{1}{t-t_k} (E_1-E_3)^\top (U_2 E_1 + U_3 E_2 + U_4 E_3) \Big\} \xi(t) \nonumber \\
    &+ 2\eta^\top(t) (\Phi_6^\top U_1\Phi_4 + U_2 E_3) \xi(t), \label{derivative_of_Uc}
    \\
    \dot{\Wc} =&~ \xi^\top(t) \Big( E_2^\top F E_2 + (\Phi_0 + \Delta_0)^\top R (\Phi_0 + \Delta_0) \nonumber \\
    &+ {\bf He} \Big\{(\Phi_0+\Delta_0)^\top(W_1(E_1-E_2)+W_2E_2 \nonumber \\
    &+ FE_1)\Big\} \Big) \xi(t) + 2\eta^\top(t)\Big(HE_1 + W_1(E_1-E_2) \nonumber \\
    &+W_2E_2 + R (\Phi_0+\Delta_0) \Big)\xi(t) + \eta^\top(t) R \eta(t). \label{derivative_of_Wc}
\end{align} \label{derivative_of_VUW}
\end{subequations}
Additionally, integrating both sides of \eqref{closed_loop_dynamic} gives
\begin{align*}
    z(t) \!- z(t_k) = (t \!-\! t_k)\Big( (\Ac \!\!+\!\! \Delta \Ac) \phi_k(t) + (\Ac_c + \Delta \Ac_c)z(t_k) \Big).
\end{align*}
Further, by multiplying both sides of the above equation with $Q\xi(t)$, we obtain
\begin{align}
    \xi^\top(t) {\bf He} \Big\{ \Phi_1^\top Q - (t-t_k)(\Phi_3 + \Delta_3)^\top Q \Big\} \xi(t) = 0. \label{equality_by_integral}
\end{align}
Therefore, from \eqref{derivative_of_VUW}, \eqref{equality_by_integral} and with the help of Lemma \ref{Lemma_integral_inequality}, we have
\begin{align}
    &\dot{V}(z) \!+ \Uc(t,z) \!-\! \Wc(t,z) \!+\! (t \!-\! t_k)\dot{\Uc}(t,z) \!+ (t_{k+1} \!- t) \dot{\Wc}(t,z) \nonumber 
    \\
    &\le \xi^\top(t) \Big( \Xi_0 + {\bf He}\big\{Y_0 \Delta_0 \big\} + (t_{k+1}-t)\big(\Xi_1 + {\bf He}\big\{ Y_1 \Delta_0 \big\} \nonumber \\
    &+ \Delta_0^\top R \Delta_0\big) + (t-t_k)\big(\Xi_2^\prime + {\bf He}\big\{ Y_2 \Delta_2 \big\}\big) - \Psi \Big) \xi(t)  \nonumber \\ &+ \eta^\top(t) \Upsilon(t)\xi(t) + \eta^\top(t) R \eta(t), \label{Lyapunov_dot}
\end{align}
where $\Xi_2^\prime = \Xi_2 + N_1^\top R^{-1} N_1 + \frac{1}{3} N_2^\top R^{-1} N_2$.

In addition, according to \eqref{bound_of_Delta_Ac} and \eqref{bound_of_Lambda_Ac_c}, one has $\|\Delta_0\|^2 = \|\Delta \Ac \|^2 + \|\Delta \Ac_c\|^2 \le 2 \delta_A^2(h,\Lambda) + \varphi^2(h,\Lambda)$; and $\|\Delta_2\|^2 = \|\Delta_5\|^2 + \|\Delta_0\|^2 + \|\Delta_3\|^2 = 3\|\Delta_0\|^2\le 3\big(2 \delta_A^2(h,\Lambda) + \varphi^2(h,\Lambda)\big)$. Thus, based on \eqref{condition_for_Lambda} and Lemma \ref{Lemma_LMIs_with_Uncertainty} with $\Gamma = \Xi_0 + h\Xi_1$, $\Pi_1 = I$, $\Delta = \Delta_0,\Pi_2 = Y_{01}^\top$, $\kappa = \frac{\varepsilon_1}{2}$, and $\Omega = h R$, the conditions \eqref{condi_2_main_theorem} and \eqref{condi_3_main_theorem} give
\begin{align}
    \Xi_0 \!+\! {\bf He}\big\{Y_0 \Delta_0 \big\} \!+\! h \big(\Xi_1 \!+\! {\bf He}\big\{ Y_1 \Delta_0 \big\} \!+\! \Delta_0^\top R \Delta_0 \big) \! \preceq \! 0. \label{a41}
\end{align}
Similarly, by using Lemma \ref{Lemma_LMIs_with_Uncertainty} with $\Pi_1 = I$, $\Pi_2 = Y_{02}^\top$, $\Delta = \Delta_2$, $\kappa = \frac{\varepsilon_2}{2}$, and $\Omega = \mathbf{0}$, and with the help of the Schur's complement, the condition \eqref{condi_4_main_theorem} yields 
\begin{align}
    \Xi_0 + {\bf He}\big\{Y_0 \Delta_0 \big\} + h\big(\Xi_2^\prime + {\bf He}\big\{ Y_2 \Delta_2 \big\}\big) \preceq 0. \label{a42}
\end{align}
Thus, from \eqref{Lyapunov_dot}-\eqref{a42}, we obtain
\begin{align*}
    &\dot{V} + \Uc - \Wc + (t-t_k)\dot{\Uc} + (t_{k+1}-t) \dot{\Wc} \nonumber
    \\
    &\le - \xi^\top(t) \Psi \xi(t) \!+\! \eta^\top(t) \Upsilon(t) \xi(t) \!+\! \eta^\top(t) R \eta(t) \nonumber 
    \\
    &\le - \lambda_{min}(\Psi) \| \xi(t) \|^2 \!+\! \varsigma \| \eta(t) \| \| \xi(t) \| + \lambda_{max}(R)  \| \eta(t) \|^2 \nonumber
    \\  
    &\le -\Big(\!\lambda_{min}(\Psi) \!-\! \frac{\varsigma s^2}{2}\!\Big) \! \|\xi(t)\|^2 \!+\! \left( \! \lambda_{max}(R) \!+\! \frac{\varsigma}{2s^2} \! \right) \! \|\eta(t)\|^2\!  \nonumber
    \\
    &\le -\Big(\!\lambda_{min}(\Psi) \!-\! \frac{\varsigma s^2}{2}\!\Big) \! \|z(t)\|^2 \!+\! \left( \! \lambda_{max}(R) \!+\! \frac{\varsigma}{2s^2} \! \right) \! \|\eta(t)\|^2\!,
\end{align*}
where $0< s < \sqrt{\frac{2\lambda_{min}(\Psi)}{\varsigma}}$ is a small positive scalar, and $\varsigma > 0$ is an upper bound of $\|\Upsilon(t)\|$ with
\begin{align*}
    \Upsilon(t) &= 2P E_1 + 2(t-t_k)(\Phi_6^\top U_1 \Phi_4 + U_2E_3) + 2(t_{k+1} - t) \\
    &\quad \times (HE_1 + W_1 (E_1-E_2) + W_2 E_2 + R(\Phi_0 + \Delta_0)).
\end{align*}
Further, the conditions \eqref{bound_of_eta} and \eqref{condition_for_Lambda_k} guarantee the bounded energy disturbance condition.
Therefore, according to Theorem \ref{theorem_stability_sampled_data_sys}, the system \eqref{closed_loop_dynamic} is IQC, then the first statement of Theorem \ref{theorem3_1} is proven. In addition, based on Theorem \ref{theorem2} with $\mu_3 =  \lambda_{min}(\Psi) - \frac{\varsigma s^2}{2} $, and $\mu_4 = \lambda_{max}(R) + \frac{\varsigma}{2s^2}$, the second statement of Theorem \ref{theorem3_2} is proven.
$\Qced$

\bibliographystyle{IEEEtran}
\balance
\bibliography{Ref}

\end{document}